\newcommand{\D}{Tajima's $D$}

\documentclass[12pt]{book}
\usepackage{fullpage}

  \usepackage{chapterbib}
\usepackage{threeparttable}
\usepackage{lscape}
\usepackage{caption}
\usepackage{subcaption}
\usepackage{graphicx}
\usepackage{amsmath}
\usepackage{longtable}
\usepackage[none]{hyphenat}
 \usepackage{setspace}
 \usepackage{url}

\usepackage{natbib}
\bibpunct[; ]{(}{)}{,}{a}{}{;}
 
 \usepackage{tocloft}

\cftpagenumbersoff{table}

\usepackage{caption}
\newcommand{\Dmel}{\emph{D. melanogaster}}

\newcommand{\Dyak}{\emph{D. yakuba}}
\newcommand{\FullDyak}{\emph{Drosophila yakuba}}

\newcommand{\Dsim}{\emph{D. simulans}}
\newcommand{\FullDsim}{\emph{Drosophila simulans}}

\newcommand{\Dere}{\emph{D. erecta}}

\newcommand{\Dros}{\emph{Drosophila}}

\newcommand{\Rebekah}{Rebek\mbox{}ah }

\usepackage{textcomp}

\begin{document}

\author{\Rebekah L. Rogers$^1$, Julie M. Cridland$^{2}$, Ling Shao$^3$, Tina T. Hu$^4$, \\ Peter Andolfatto$^4$, and Kevin R. Thornton$^3$}

\title{Tandem duplications and the limits of natural selection in \FullDyak {} and \FullDsim }
\date{}

{\hfill \textbf{}
\let\newpage\relax\maketitle}

\begin{center} \Large Research Article \end{center}
\begin{center}Biological Sciences, Poulation Biology\end{center}
\vspace{0.25in}

\noindent 1) Ecology and Evolutionary Biology, University of California, Berkeley \\
\noindent 2) Ecology and Evolutionary Biology, University of California, Davis \\
\noindent 3) Ecology and Evolutionary Biology, University of California, Irvine \\
\noindent 4) Ecology and Evolutionary Biology and the Lewis Sigler Institute for Integrative Genomics, Princeton University \\

\noindent \textbf{Running head: } Tandem duplications in non-model \Dros

\vspace{0.25in}


\vspace{0.25in}

\noindent \textbf{Corresponding author:} Rebekah L. Rogers,  Dept. of Ecology and Evolutionary Biology, 5323 McGaugh Hall, University of California, Irvine, CA 92697 \\
\\
\noindent \textbf{Phone:}  949-824-0614

\noindent \textbf{Fax:}  949-824-2181

\noindent \textbf{Email:} rogersrl@uci.edu \\

{\setlength{\baselineskip}
{2.0\baselineskip}
\renewcommand{\baselinestretch}{2.0}

\newpage 
\section*{Abstract}
{\setlength{\baselineskip}
{2.0\baselineskip}
\renewcommand{\baselinestretch}{2.0}
Tandem duplications are an essential source of genetic novelty, and their variation in natural populations is expected to influence adaptive walks. Here, we describe evolutionary impacts of recently-derived, segregating tandem duplications in \FullDyak {} and \FullDsim. We observe an excess of duplicated genes involved in defense against pathogens, insecticide resistance, chorion development, cuticular peptides, and lipases or endopeptidases associated with the accessory glands, suggesting that duplications function in Red Queen dynamics and rapid evolution. We document evidence of widespread selection on the \Dsim {} X, suggesting adaptation through duplication is common on the X. Despite the evidence for positive selection, duplicates display an excess of low frequency variants consistent with largely detrimental impacts, limiting the variation that can effectively facilitate adaptation. Although we observe hundreds of gene duplications, we show that segregating variation is insufficient to provide duplicate copies of the entire genome, and the number of duplications in the population spans 13.4\% of major chromosome arms in D. yakuba and 9.7\% in D. simulans. Whole gene duplication rates are low at  $1.17\times10^{-9}$ per gene per generation in \Dyak {} and $6.03\times10^{-10}$ per gene per generation in \Dsim, suggesting long wait times for new mutations on the order of thousands of years for the establishment of sweeps. Hence, in cases where adaption depends on individual tandem duplications, evolution will be severely limited by mutation. We observe low levels of parallel recruitment of the same duplicated gene in different species, suggesting that the span of standing variation will define evolutionary outcomes in spite of convergence across  gene ontologies consistent with rapidly evolving phenotypes.}
}
\chapter*{}
{\setlength{\baselineskip}
{2.0\baselineskip}
\renewcommand{\baselinestretch}{2.0}

\section*{Introduction}

Tandem duplications are an essential source of genetic novelty that is useful for the development of novel traits \citep{WolfeReview, Ohno1970} and their prevalence in populations is therefore expected to influence the arc of evolutionary trajectories.  The observed landscape of tandem duplications in \Dros {} spans only a few percent of the genome \citep{Rogers2014, Emerson2008, Cardoso2011, Dopman}, and it is unclear to what extent duplications, whether from newly arising or from standing variation, can provide a sufficient source of adaptive genetic variation.  Tandem duplications produce a variety of novel gene structures including chimeric genes, recruited non-coding sequence, dual promoter genes, and whole gene duplications \citep{Rogers2014,QZhou2008, Katju2006}.  Surveys within single genomes have suggested that whole gene duplications may form at low rates in comparison with SNPs, with even lower mutation rates for complex variants such as chimeric genes \citep{RBH, Rogers2014, QZhou2008}.  Yet, these alternative genetic structures are known forces of evolutionary innovation.   Whole gene duplications often develop novel functions or specialize in ancestral functions \citep{WolfeReview}, and chimeric genes are more likely still to produce novel molecular effects and play a role in adaptive evolution \citep{RapidEvolution}.  Although these variants contribute substantially to the evolution of genome content \citep{QZhou2008, RBH, Lynch2000}, their lower rates of formation may render evolution of tandem duplications more likely to be limited by mutation.  

 If population-level mutation rates are sufficiently large, new mutations will accumulate quickly and adaptation is expected to proceed rapidly \citep{Hermisson2005}.   However, if population-level mutation rates are low, then there will be long wait times until the next new mutation and evolutionary trajectories are likely to stall at suboptimal solutions during the mutational lag \citep{Maynard1971,Gillespie1991,Hermisson2005}.  \Dros {} have large population sizes in comparison to other multicellular eukaryotes with $N_e\approx 10^5-10^6$ \citep{Kreitman1983, Bachtrog2006,AndolfattoNe} and absolute numbers of individuals large enough to provide large numbers of SNPs at many sites every generation \citep{Karasov2010}.  However, the prevalence of other types of mutations beyond SNPs has not been systematically surveyed.  If the supply of tandem duplications is limited by mutation, we expect to see suboptimal outcomes in adaptive walks, limited ability to adapt to changing environments, and low rates of evolution through parallel recruitment of the same genetic solutions in different species.   The \Dros {} offer an excellent model system for population genomics, allowing for a whole genome survey of the genetic landscape of standing variation across species in natural populations and determination of genetic convergence across taxa.  There are multiple sequenced reference genomes for \Dros, and genomes are small and compact, allowing for whole genome population surveys using next generation sequencing.  Here, we focus on \Dyak {} and \Dsim, which are separated by 12 MY of divergence \citep{Tamura2004}, allowing for surveys of distantly related groups which are not expected to share polymorphic variation due to ancestry.  Thus, we can measure the limits of standing variation and the incidence of parallel duplication across species, which should be broadly applicable to multicellular eukaryotic evolution.

Convergent evolution is regarded as the ultimate signal of natural selection: if the same solution is favored for a given environment then selection should result in similar phenotypes \citep{Gould1979}.   There are many known cases of convergent phenotypic evolution, but the understanding of convergence at the genetic level is limited to a small number of case studies across diverse clades \citep{Stern2013}. These case studies have revealed convergent evolution through different genetic solutions in vertebrates \citep{Chen1997, Shapiro2009, Brodie2010}, and arthropods \citep{Khadjeh2012, Wittkopp2003, Tanaka2009, Zhen2012}.   Parallel evolution through similar genetic solutions, however, appears to be more common at mutational hotspots where high mutation rates at targeted sites produce mutations at a steady rate \citep{Riehle2001,Chan2010, Moxon1998}.   Beyond these results from natural populations, convergence has often been observed in experimental evolution and is considered a signal of selection favoring alleles \citep{Riehle2001, Moxon1998, Burke2010,Woods2006, Tenaillon2012}. However, most studies of laboratory evolution take advantage of microbes or viruses with large population sizes roughly $10^{9}-10^{10}$ such that every mutation is likely to be sampled every generation \citep{Riehle2001,Moxon1998} or from small populations that share a common pool of standing variation \citep{Burke2010, Orozco2012} and may therefore be qualitatively different outcomes in comparison to natural evolution in multicellular eukaryotes.  Indeed, known examples of evolution through parallel recruitment of the same genetic solutions in natural populations often occur through a common ancestral genetic pool \citep{Colosimo2005} or through introgression \citep{Martin2012}.  

The instance of convergent evolution across unrelated taxa that do not share ancestry is essential to understanding the ways mutation limits evolution, the role of standing variation in evolutionary trajectories, and the genetic architecture of adaptation.   Here, we survey standing variation for tandem duplications in \FullDyak {} and \FullDsim {} and the role that this standing variation plays in adaptive evolution in natural populations.  We identify an overrepresentation of tandem duplications with gene ontologies consistent with rapidly evolving phenotypes, signals of reduced diversity surrounding tandem duplications, and an overabundance of high frequency variants on the \Dsim {} X chromosome, pointing to a role for adaptation through gene duplication.  We further show that the span of tandem duplications in populations is limited to a small fraction of the genome and that low mutation rates will lead to long wait times for sweeps on new mutations. These results imply that evolution by tandem duplication will be limited by mutation and that parallel recruitment of gene duplicates across species is likely to be exceedingly rare even in the face of strong selection on similar phenotypes in different species.  


\section*{Results}
We previously identified hundreds to thousands of segregating duplications in natural populations of \Dyak {} and \Dsim, including large numbers of gene duplications (Table \ref{GeneDups}) \citep{Rogers2014}. We assess the numbers and types of gene duplications, differences in duplication rates across species and explore the limits of the landscape of standing variation for tandem duplications present in each species to determine the extent to which these variants can serve as a source of genetic novelty. 

Recently derived, segregating tandem duplications were detected using paired end reads and coverage changes in \Dyak {} and \Dsim {} in samples of 20 isofemale lines derived from natural populations of each species \citep{Rogers2014}.  Using divergently oriented paired-end reads, we identified 1415 segregating tandem duplications in \Dyak, in comparison to 975  in \Dsim.  The tandem duplications identified across these sample strains cover 2.6\% of assayable the genome of the X and 4 major autosomes in \Dyak {} and 1.8\% of the assayable genome of the X and 4 major autosomes in \Dsim.  These variants have a high (96\%) confirmation rate \citep{Rogers2014} and constitute a high quality dataset for population genomics.

\subsection*{Widespread selection on the \Dsim {} X chromosome}

If tandem duplications are common targets for adaptation and selective sweeps, we should observe a shift in the site frequency spectrum (SFS) toward high frequency variants relative to neutral markers \citep{Nielsen2005}.  We compare the SFS for duplications with the SFS for SNPs from 8-30 bp of short introns used as a putatively neutral proxy to determine whether duplicates are subject to selection (Figure \ref{SFSSNPComp}). The SFS for duplications is significantly different from that of intronic SNPs on the \Dsim {}  autosomes ($W = 268$, $P= 2.981\times10^{-6}$) and \Dyak {} autosomes ($W = 212$, $P = 3.507\times10^{-6}$).  In \Dyak {} the SFS for duplicates on the X is significantly different from that of SNPs ($W = 211$, $P = 4.781\times10^{-4}$) using a Wilcoxon sign rank test.  Duplicates show an excess of singleton variants on the autosomes in both species (Figure \ref{SFSSNPComp}), suggesting deleterious impacts on average.    We find a significant difference between the SFS of duplicates on the X chromosome and the autosomes in \Dyak {} ($W=172$, $P=0.0128$) but not in \Dsim {} ($W=183.5$, $P=0.1848$) (Figure \ref{SFSSingleSm}, Table \ref{SFSKruskaltab}-\ref{SFSKolmogorovtab}).  In contrast, the SFS for duplicates on the D. simulans X chromosome exhibit fewer singletons in comparison to neutral SNPs and a peak of high frequency duplicates (Figure \ref{SFSSingleSm}).  We observe an excess of duplications at a sample frequency of 20 out of 20 sample strains (with no indication of duplication or misassembly in the resequenced reference) on the X chromosome of \Dsim {} in comparison to neutral SNPs ($P<10^{-6}$), but observe no duplicates at a sample frequency of 20 out of 20 in \Dyak {} on the X or autosomes.  Furthermore comparisons of the SFS show an excess of highest frequency variants $\geq$16 out of 17 on the X ($\chi^2 = 21.8334$, $df = 1$, $P = 2.974\times10^{-6}$).    The excess of high frequency duplicates on the \Dsim {} X chromosome is indicative of selection favoring large numbers of tandem duplicates.  These results imply that adaptation through duplication is common on the \Dsim {} X.  
 
 We have calculated average heterozygosity per site ($\theta_{\pi}$) \citep{Tajima1983}, Wattersons's $\theta$ ($\frac{S}{a}$ per site) \citep{Watterson1975}, and \D {} \citep{Tajima1989} for the four major autosomes and the X chromosome in \Dyak {} and \Dsim {} using 5 kb windows with a 500 bp slide correcting for the number of sites with coverage sufficient to confidently identify SNPs  (Figure \ref{ChromPlotsDyak2L}-\ref{ChromPlotsDsimX}).    We compare $\theta_{\pi}$ in regions surrounding tandem duplications and putatively neutral SNPs from 8-30 bp of short introns to search for signals of reduced diversity consistent with selection acting on tandem duplications. These tandem duplications are polymorphic and represent putative sweeps in progress and such comparisons to within-genome controls of neutral SNPs offer greater power than alternative tests of selection \citep{Coop2012}.   We compare estimates of $\theta_{\pi}$ for windows from 50 kb upstream to 50kb downstream of a duplicated locus with comparable windows surrounding derived  segregating SNPs from putatively neutral sites of short introns, excluding centromeric regions.  We find a reduction in $\theta_{\pi}$ per site surrounding newly arisen duplications on \Dyak {} chromosome 3R (single tailed Wilcox test, $P=0.0064$, see Table \ref{ReducedDiversityTab}, Figure \ref{DistanceFromDyak}) as well as on \Dsim {} chromosomes 2L, 3L, 3R, and the \Dsim {} X (single tailed Wilcox test, $P \leq 0.05$; see Table \ref{ReducedDiversityTab}) (Figure \ref{DistanceFromDsim}).  Chromosome 3L has reduced diversity surrounding duplicates in \Dsim, even though a region encompassing multiple duplications with signals of a broad selective sweep encompassing multiple loci at roughly 8.5 Mb is excluded (Figure \ref{ChromPlotsDsim3L}).  A significant excess of diversity is seen on chromosome 2L in \Dyak {} ($W = 31594$, $P = 2.517\times10^{-4}$), a possible product of alternative evolutionary dynamics driven by segregating inversions on 2L  \citep{Lemeunier1976,Llopart2005}.   We further compare  $\theta_{\pi}$ for windows centered about duplications that capture solely intergenic sequence with those that capture coding sequence but respect gene boundaries, and those that create chimeric genes or recruited non-coding sequence.   In \Dsim, gene duplications that do not create chimeric genes have reduced diversity  ($W = 17880$, $P = 0.01233$) in comparison to mutations that capture intergenic mutations (Figure \ref{ThreeTypes}C-D), but chimeric gene duplications do not ($W = 5020$, $P= 0.5755$), consistent with selection driving an excess of whole gene duplications in \Dsim {} \citep{Rogers2014}, but relationships in \Dyak {} are not signficant for chimeric gene mutations ($W = 2656$,  $P=0.8226$) or whole gene duplications ($W = 6387$,$P = 0.8847$).  All of these mutations are polymorphic indicating sweeps in progress or neutral dynamics and power may be limited to detect differential selection on different types of mutations.  Based on a binomial test, we observe a marginally significant overrepresentation of duplications that capture gene sequences vs solely non-coding sequences in \Dyak {} ($P$=0.029) and highly significant for \Dsim {} ($P=9.044\times10^{-5}$). 
 
 While demography and neutral evolutionary forces can result in shifts of diversity and site frequency spectra, these forces should affect sequences across individual chromosome arms and act similarly on intronic SNPs, and are therefore unlikely to explain the observed differences between duplicates and intronic SNPs.  Therefore, it seems likely that the overabundance of high frequency variants on the \Dsim {} X is driven by natural selection.  Thus, we would expect many of these high frequency variants to be strong candidates for ongoing selective sweeps.  We further observe large numbers of singleton variants among tandem duplicates in comparison with intronic SNPs in \Dyak {} and \Dsim {} autosomes.  Copy number variants are subject to purifying selection in \Dmel {} \citep{Emerson2008, JCridland}, and we observe large numbers of singleton variants in excess of neutral expectations, indicating negative selection preventing variants from rising to higher frequency.  Hence, while some variants are likely to offer a means of adaptive change, many are likely to ultimately be lost from the pool of standing variation.  We suggest that tandem duplications are likely to confer phenotypic impacts that are on average large enough to surpass the threshold of near neutral effects.

\subsection*{Rapid Evolution}
Biases in the rates at which duplications form in different genomic regions or a greater propensity for selection to favor duplications in specific functional classes can result in a bias in gene ontology categories among duplicated genes.  We use DAVID gene ontology analysis software to identify overrepresented functions among duplicate genes in \Dyak {} and \Dsim {} \citep{Rogers2014}.  While some categories are specific to each species, immune response and toxin metabolism are overrepresented in both species (Figure \ref{VennDual}A,Table \ref{GOBiasDual}), functions that have also been identified as being overrepresented in a parallel study of \Dmel {} \citep{Zichner2013}.  We also observe an overrepresentation of chitin cuticle formation and chemosensation in both \Dyak {} and \Dsim{} (Table \ref{GOBiasDual}). Furthermore, lipases and endopeptidases commonly found in the accessory glands are present among high frequency variants in both species and there are multiple independent duplications in chorion and egg development genes (Table \ref{GOBiasDual}), suggesting a role for duplications in sexual conflict. The overabundance of toxin metabolism genes and immune response peptides in both species as well as the overrepresentation of putative chemoreceptors, chitin based cuticle genes, endopeptidases, and oogenesis factors suggests that duplications may be key players in rapidly evolving systems.  Moreover, the strong agreement in overrepresented functional categories points to strong selective pressures acting in parallel in these independently evolving species.  To determine whether selection is favoring these functional classes, we identified duplications outside centromeric regions that lie in windows at or below the 5\% tail of $\theta_{\pi}$, consistent with selection reducing diversity.  Among these genes in regions with reduced diversity we identify genes in both \Dyak {} and \Dsim {} with functions in chorion or oogenesis, mating behavior, immune response and defense against bacteria, olfactory response, chitin metabolism, xenobiotics and toxin metabolism, and sperm development (Supplementary Information).  The presence of genes with these functional categories is consistent with a portion of the overrepresentation in gene ontologies across all duplicates being driven at least in part by selection.

 \subsection*{Limits of standing variation in natural populations}

We observe hundreds of segregating tandem duplicates in \Dyak {} and   \Dsim, spanning  2.6\% of assayable the genome of the X and 4 major autosomes in \Dyak {} and 1.8\% of the assayable genome of the X and 4 major autosomes in \Dsim.  If evolutionary trajectories depend on duplications to effect beneficial phenotypic changes, then the number of segregating variants available in the population may not contain desired variants as standing variation.  We estimate the number of variants present in the entire population based on the observed sample variation in order to determine the extent to which selection will be limited by mutation. We estimate that the population contains at most  7,230 segregating tandem duplications in \Dyak {} and 4,720 in \Dsim {} (Table \ref{SegSites}-\ref{Tiled}), corresponding to 13.4\% of major chromosome arms in \Dyak {} and 9.7\% of major chromosome arms in \Dsim.  Thus, the standing variation for tandem duplications will be insufficient to offer tandem duplications for every potential gene across the entire genome and for the majority of the genome ($\approx$85\%), if a tandem duplication is required for adaptation, evolutionary trajectories must by definition rely on new mutations. 

We calculate population level mutation rates $\theta_{\pi}$ ($4N_e \mu$) of 0.00277 for whole gene duplications  0.00082 for recruited non-coding sequence and 0.00088 for chimeras in \Dyak.  Population level mutation rates in \Dsim {} are  slightly higher with 0.00291 for whole gene duplications  0.00117 for recruited non-coding sequence and 0.00041 for chimeras.  In comparison, we calculate $\theta_{\pi}$ for putatively neutral intronic SNPs of 0.0138 for \Dyak {} and 0.0280 for \Dsim.  We use these estimates of $\theta$ to calculate the likelihood of adaptation from alleles among the standing variation rather than new mutation for a population ($P_{sgv}$)   \citep{Hermisson2005}  assuming additive variants with a large selection coefficient of 1\% under an additive genetic model.   With such low levels of $\theta_{\pi}$ the likelihood of adaptation from a tandem duplication among the standing variation is 2.2\% in \Dyak {} and 2.6\% in \Dsim {} (Table \ref{MutRates}), a strikingly low likelihood that standing variation offers a sufficient substrate for adaptation. Even with a massive selective coefficient of $s=0.20$ \citep{Karasov2010}, the likelihood of adaptation from standing variation rather than new mutation is 3.1\% for duplicates in \Dyak {} and 3.4\% in \Dsim.   Chimeras are even more extreme with less than a 1\% chance of fixation from standing variation (Table \ref{MutRates}). In comparison, intronic SNPs have a likelihood of adaptation from standing variation of 12.1\% in \Dyak {} and  24.6\% in \Dsim {} given s=0.01, and 15.7\% in \Dyak {} and 30.1\% in \Dsim, given extreme selection coefficients of s=0.20 (Table \ref{MutRates}).  Thus, the limits of standing variation are expected to be far more severe for complex gene structures than for SNPs.  

 We calculate the per generation mutation rate $\mu$ per gene for whole gene duplications, considering duplicates that capture 90\% or more of gene sequences, in agreement with previous methods \citep{RBH}. We estimate a whole gene duplication rate of $1.17 \times 10^{-9}$ per gene per generation for \Dyak {}  and $6.03 \times 10^{-10}$ per gene per generation for \Dsim {} (Figure \ref{PhyloNums}, Table \ref{MutRates}), in general agreement with estimates derived from surveys of duplicates in the \Dmel {} reference genome of $3.68 \times 10^{-10}$ per gene per generation \citep{RBH, QZhou2008}.   The rate of recruited non-coding sequence is $3.46\times10^{-10}$ in \Dyak {} and $2.42\times10^{-10}$ in \Dsim {} and the rate of chimeric gene formation is equally low with $3.7\times10^{-10}$ in \Dyak {} and $8.52\times10^{-11}$ in \Dsim {} (Figure \ref{PhyloNums}, Table \ref{MutRates}).   We observe more duplications in \Dyak {} in spite of its lower $N_e$, yielding a duplication rate per gene in \Dyak {} two-fold higher than that of \Dsim.  Given these estimates of $\theta_{\pi}$, we estimate $T_e$, the time to establishment of a deterministic sweep from new mutations in a population such that variants overcome the forces of drift \citep{Gillespie1991,Maynard1971}, assuming that beneficial mutations appear at strongly selected sites at a rate equivalent to the genome-wide mutation rate.  In reality not all mutations are beneficial and the rate of adaptive substitution is likely to be less common. These estimates therefore represent an lower bound on the time to adaptation through new mutation.   With a small selection coefficient of s=0.01, in \Dyak {} $T_e$ would be 7270 generations (~600 years at 12 generations per year) for SNPs, 36,000 generations (3000 years) for whole gene duplications, and over 100,000 generations ($\geq$ 9500 years) for chimeric genes (Table \ref{MutRates}).   For \Dsim, these numbers point to a greater disparity between SNPs and duplications with $T_e$ of 3580 generations (~300 years) for SNPs, 34,400 generations (2800 years) for whole gene duplications, and 243,000 generations (20,000 years) for chimeric genes (Table \ref{MutRates}).  Under extreme selection coefficients and with the assumption that the beneficial mutation rate matches the mutation rate per site, wait times may be shorter, allowing for adaptation at SNPs in hundreds of generations (decades) and thousands of generations (centuries) for duplications (Table \ref{MutRates}), though such extreme dynamics are unlikely to reflect the range of selection coefficients or the rate of adaptation for the genomewide \citep{Jensen2008,Andolfatto2007}. These estimates of mutation rates for whole gene duplications and complex gene structures point to long wait times for new mutations and a disparity in the response of duplicates and SNPs in the face of strong selective pressures.  Although the differences in mutation rates appear to be modest, they can result in additional thousands of years in the wait time for selective sweeps to establish with new mutations, resulting in limited ability to adapt to shifting selective pressures.
 
A Bayesian binomial estimate of the 95\% lower CI suggests that for all mutations not observed across the 20 sample strains, their frequency in the population is $\leq 0.1329 $, with a 50\% lower CI of $\leq 0.0325$.  Very rare mutations may have difficulty escaping the forces of drift in the population, especially if recessive \citep{Haldane}, and therefore ultimately the number of duplications that are at frequencies high enough to establish deterministic selective sweeps will be considerably fewer that than the number that exist in the population.  The number of tandem duplications that have the potential to sweep to fixation may be substantially less than indicated by the number of segregating sites.  Thus, the pool of standing variation in tandem duplications will provide only a limited substrate of novel genetic sequences and evolution will be limited by mutation.

Some 56 genes are partially or wholly duplicated both in \Dyak {} and in \Dsim, a mere 11\% of genes duplicated genes in \Dsim {} ($\frac{56}{478}$) and less than the number of genes duplicated multiple times in \Dsim {} alone, suggesting that there is little concurrence in the standing variation of the two species.  That 56 genes would be shared across the two species is greater than expected given the limits of available standing variation of 478 duplicated genes in \Dsim {} and 875 in \Dyak {} based on uniform chance ($P= 2.812\times10^{-8}$, binomial test) pointing to mutational or selective pressures on similar genes (SI Text).   \Dsim {} shares fewer genes with \Dmel {} than does \Dyak, a result of fewer annotated gene models in \Dsim {} \emph{w501} reference \citep{SimRef}.   Furthermore, a comparison to duplicate genes in \Dmel {} \citep{Zichner2013} shows only 5 genes that exist among the segregating variation of tandem duplications in all three species.  The mutations described here have been polarized with respect to ancestry, and are segregating meaning that they are expected to have formed very recently.  As such, shared variants are the product of independent mutation in the two species, not shared ancestry.   We find that 13.4\% of the genome is present but unsampled in \Dyak {} and 9.7\% in \Dsim, indicating that the likelihood of shared, unsampled variation is low.  Such unsampled alleles will be at low frequency and are unlikely to be able to establish selective sweeps. Hence, the portion of variation available for selective sweeps that is shared across species will be low, resulting in a rarity of evolution through parallel recruitment of tandem duplicates.   Some genes within the genome are captured by as many as 6 independent duplications in \Dyak {} and 32 independent duplications in \Dsim.  There are 10 genes in \Dyak {} and 12 genes in \Dsim {} that are captured by more than three independent duplications, and these have been excluded from mutation estimates (Table \ref{MultipleDups}).  These ``hotspots" within the genome may have duplication rates high enough that evolution will not be subject to the same limitations with respect to standing variation and sweeps on new mutations.  

\section*{Discussion}
We have described the prevalence of tandem duplications in natural populations of \Dyak {} and \Dsim, their frequencies in the population, and the genes that they affect. We find that duplications show a bias towards gene ontologies associated with rapid evolutionary processes and that they commonly affect the X chromosome in \Dsim {} in comparison to the autosomes.  In spite of their strong role in adaptation, we find low rates of parallel recruitment of tandem duplications across species due to low formation rates and mutation limited evolution.

\subsection*{Widespread positive selection on the X chromosome in \Dsim}

We observe an excess of high frequency duplications on the \Dsim {} X chromosomes in comparison to neutral intronic SNPs as well as sings of reduced diversity surrounding duplications on the \Dsim {} X, consistent with widespread selection.  Background selection \citep{Charlesworth1993} and hitchhiking \citep{MaynardSmith1974} are not expected to act differently on duplications in comparison to SNPs and cannot explain the patterns observed.  Yet, we observe significant differences between the SFS of duplicates and putatively neutral SNPs, pointing to a role for adaptation through tandem duplication.  We also observe reduce nucleotide diversity surrounding tandem duplications on the \Dsim {} X, consitent with selection favoring duplicates.  Hence, the overabundance of high-frequency duplications on the X is likely to be driven by selection and these represent strong candidate loci for ongoing selective sweeps.  Based on the newly assembled \Dsim {} reference, X vs. autosome divergence indicates faster evolution on the X chromosome at non-synonymous sites, long introns, and UTRs \citep{SimRef}.  This pattern is distinct from observations at synonymous sites as well as general patterns of differential evolution on the autosomes \citep{SimRef}, further evidence of more frequent selective sweeps on the X chromosome.    

The X chromosome is thought to evolve rapidly due to sexual conflict, intragenomic conflict, and sexual selection \citep{Presgraves2008} and thus multiple selective forces may facilitate the spread of duplicates on the X.  The X chromosome in \Dsim {} houses an excess of duplicates in comparison to all autosomes, as well as a strong association with repetitive sequence and tandem duplications on the X \citep{Rogers2014}.  Therefore, the X chromosome appears to be subject to particularly rapid evolution in duplicate content in \Dsim.   Previous work has identified signals of adaptation through duplication on the \Dmel {} X chromosome \citep{Thornton2002, Thornton2005}, suggesting parallel evolution through duplication in these species.  However, we do not observe similar patterns in \Dyak, suggesting that the X may either be evolving under different selective pressures in the two different clades or that selective pressures on the \Dyak {} X chromosome are of lesser magnitude.  Stronger sexual selection, greater selection for X-chromosome related traits, sympatric associations with competitor species with reinforcement for mating aversion, or a greater instance of driving X chromosomes might potentially drive these species differences in X-chromosome evolution, and elucidating the nature of selection on these sex chromosomes may help explain the adaptive (or selfish) role of tandem duplication on the X. 
 
\subsection*{Mutation limited evolution}

While both \Dsim {} and \Dyak {} house a rich diversity of duplicated sequences, only a few percent of the genome will be covered by tandem duplications.   With lower mutation rates for duplications \citep{Lynch2003, RBH, QZhou2008}, there may be long wait times to achieve any single new mutation, and the landscape of standing variation will shape evolutionary outcomes.  As such, any evolutionary path that is dependent upon duplications of any specific genomic sequence will be severely limited by the small likelihood that the necessary mutation is among the standing variation in tandem duplications.  \Dros {} represent organisms with large effective population sizes (Figure \ref{PhyloNums}) \citep{Bachtrog2006, EyreWalker2002} and are expected to host large numbers of duplications as standing variation in comparison to other multicellular eukaryotes.  We have shown that the number of tandem duplications segregating in the population is substantially smaller than the number of mutations needed to guarantee a duplicate of any desired genomic region.   However, when population level mutation rates are small, standing variation is unlikely to offer a sufficient substrate for selective sweeps and systems will be stuck waiting for new mutations that are slow to materialize \citep{Hermisson2005}.  We observe population level mutation rates $\theta$ per gene for tandem duplications on the order of 0.00277 in \Dyak {} and 0.00291 in \Dsim {} (Table \ref{MutRates}, Figure \ref{PhyloNums}) resulting in low probabilities that standing variation offers the major source of adaptation and long wait times to sweeps on new mutations on the order of hundreds to thousands of years.  While retrogenes might provide additional sources of duplicated sequences, their rates of formation are exceptionally limited \citep{Schrider2011,QZhou2008} and they are therefore not expected to contribute more substantially than tandem duplications to genomic variation and will not suffice to overcome these limitations of low mutation rates.  Thus, we conclude that outside of a small number of mutational hotspots evolution through duplication is mutation limited even in \Dros {} which have large $N_e$, and that these limits are expected to be even more severe for many other multicellular eukaryotes, especially vertebrates.  

The majority of tandem duplications identified in \Dyak {} and \Dsim {} appear to be at extremely low frequency, with an excess of singleton variants in comparison to neutral intronic SNPs, suggesting that large numbers of duplications are detrimental, consistent with previous work in other species \citep{Emerson2008}.  It has previously been argued that the accumulation of duplications is the product of small $N_e$ and inability of selection to purge nearly neutral alleles from the population \citep{Lynch2000, LynchBook}.  However, we show that duplicates are less likely to be neutral in comparison to putatively neutral intronic SNPs suggesting that they often have phenotypic effects larger than the limit near-neutrality.  We have shown that both positive and negative selection will affect the fixation or loss of duplications and that simplified nearly neutral theories are unlikely to explain the patterns observed across species.  Rather, selection is expected to play an appreciable role in the evolution of tandem duplications and their contribution to genome content.    

\subsection*{Likelihood of parallel recruitment of tandem duplications across species}

Convergent evolution is often interpreted of a signal of adaptation in experimental evolution and in natural populations \citep{Gould1979,Stern2013}. Here, we show that for tandem duplications, parallel recruitment of genes for duplication and diversification independent from shared ancestry will be very rare in spite of convergence in functional categories represented.  Thus, the reliance on genetic convergence to establish natural selection in natural populations will underreport selected alleles and result in significant underestimation of the number and types of alleles that are selected.  Though convergence is common in experimental evolution of both prokaryotic systems and  multicellular eukaryotes with shared ancestry, these results suggest that such convergence is unlikely to reflect the frequency of convergent evolution in natural populations of independently evolving species of multicellular eukaryotes that have little shared standing variation.   We observe an excess of variants with gene ontologies consistent with similar rapid evolutionary processes both in \Dyak {} and in \Dsim {} (Figure \ref{VennDual}A).  However, few genes ($\sim 11\%$) are duplicated in both species and only a handful have been identified in \Dsim, \Dyak, and \Dmel {} (Figure \ref{VennDual}B).  Moreover, none of the high frequency variants in the in \Dyak {} and \Dsim {} capture orthologous sequences.  Hence, in spite of parallel selective pressures on rapidly evolving phenotypes, there is little parallel recruitment of the same genetic solutions l with respect to duplication.  Given the limited genomic span of standing variation in the population (Table \ref{Tiled}), and low rates of new mutation (Figure \ref{PhyloNums}, Table \ref{MutRates}), as well as the low frequency of a large fraction of variants, parallel fixation of tandem duplications in the same genes will be extremely rare even among genera with large effective population sizes facing similar selective pressures.  

Convergence depends on the wait time of new mutations to enter populations and establish selective sweeps.  We show that the average wait time for a new mutation given a selection coefficient of $s=0.01$ is hundreds of years for SNPs.  Here, we find that tandem duplications display signals of reduced heterozygosity in the surrounding regions as well as an association with gene ontologies indicative of rapidly evolving phenotypes, and an overrepresentation of shared tandem duplicates across species for specific genes given the limits of standing variation, consistent with widespread adaptation through tandem duplication.  However, the average wait time for a deterministic sweep to establish in a population will be thousands of years for tandem duplications and tens of thousands of years for chimeric genes given a modest selection coefficient of $s=0.01$.  Such strongly selected sites are expected to be rare throughout the genome and beneficial mutations are likely to appear less often than the actual mutation rate  \citep{Jensen2008,Andolfatto2007}.  Thus, these wait times given strong selection provide a lower bound to the wait time for a selected sweep.    We therefore expect that mutation will severely limit evolution through whole gene duplication and chimera formation, to the extent that adaptation depends on tandem duplications, the ability of organisms to adapt to changing environments will be hindered by a lack of variation.  Thus, even when a given tandem duplication is needed for adaptation, we expect that the limits of mutation will lead to low levels of convergence and scarcity of shared genetic solutions. 

\subsection*{Duplicate genes and rapidly evolving phenotypes}
Both \Dsim {} and \Dyak {} have an overabundance of genes with ontology classifications involved in immune function, chemosensory processing or response, and drug and toxin metabolism (Table \ref{GOBiasDual}).   Furthermore the instance of independent duplications confirm a bias toward chemosensory receptors, chorion development and oogenesis, as well as immune response \citep{Rogers2014}.  These phenotypes are strongly associated with rapid evolution due to host-parasite interactions, predator-prey coevolution, and sexual conflict \citep{Lazarro2012,Beckerman2013,InsecticideReview2,Panhuis2006}.  Previous work has observed similar bias toward rapid amino acid substitutions in olfactory genes, and chitin cuticle genes in \Dmel {} and \Dsim {}  \citep{Begun2007}, selection for gene family evolution in and selection for toxin resistance is common in \Dmel {} \citep{TEInsecticide,InsecticideReview2} suggesting that associated phenotypes may be under widespread selection in multiple species.

Host pathogen systems as well as arms races in pesticide and toxin resistance operate under Red Queen dynamics in which conflicts between organisms result in repeated selective sweeps \citep{vanValen}.  Organisms that lack the genetic means to adapt to rapidly changing systems will be at a distinct disadvantage in the face of selective events.   Additionally, the overrepresentation of duplications in cytochromes and drug or toxin metabolism genes confirms rapid evolution in copy number seen in comparison of reference genomes \citep{TwelveGenomes} as well as recent studies of insecticide resistance and viral resistance in natural populations \citep{TEInsecticide, InsecticideReview, Magwire2011}.   Large amounts of divergence driven by selection among non-synonymous sites and UTRs in \Dsim {} \citep{Haddrill2008} and high rates of adaptive substitutions \citep{Andolfatto2011, Begun2007} point to widespread selective pressures acting in \Dsim, and it is likely that these same pressures influence the current diversity and frequency of copy number variants.    If rapidly evolving systems rely heavily on complex mutations or if selection coefficients are modest, profiles of standing variation will place strong limits on outcomes in response to selection.

Shifting selective pressures such as those found in rapidly evolving systems or gross ecological change require a pool of genetic variation to facilitate adaptation.  We observe standing variation and mutational profiles that will limit evolutionary trajectories and would expect these limits to be even more severe for rapidly evolving phenotypes.   Repeated sweeps are expected to purge genetic and phenotypic diversity, and recovering such diversity after sweeps can take thousands of generations \citep{Kaplan1989}.  Thus, during rapid evolution, selection will potentially purge diversity that is needed for subsequent steps in the adaptive walk.  Hence, although duplications are key players in rapid evolution, their limited rates of formation combined with low frequencies due to commonly detrimental impacts will hinder evolutionary outcomes or force alternative adaptive trajectories precisely when variation is urgently needed.  Moreover, large numbers of duplicates are low-frequency, suggesting that detrimental impacts further limit standing variation.  Thus, we conclude that the available substrate of tandem duplications and profiles of standing variation will define evolutionary outcomes in \Dros {} and other multicellular eukaryotes.

\section*{Materials and Methods}
\subsection*{Tandem duplications}
Tandem duplications were identified using paired-end Illumina sequencing of genomic DNA for 20 strains of \Dyak {} and 20 strains of \Dsim {} as well as the reference genome of each species as described in Rogers et al. 2014.  The dataset describes derived, segregating tandem duplications that span 25 kb or less.   These sequences exclude ancestral duplications as well as putative duplications in the resequenced reference genomes.  The resulting list of variants describes segregating variation for newly formed tandem duplicates across the full genome in these two species of non-model \Dros.   

\subsection*{Identifying duplicated coding sequence} 

Tandem duplications were previously identified using a combination of paired end read mapping and coverage changes in 20 isofemale lines of \Dyak {} and 20 isofemale lines in \Dsim {} generated via 9-12 generations of sibling mating from wild-caught flies.  We sequenced 10 isofemale lines of \Dyak {} from Nairobi, Kenya, and 10 isofemale lines from Nguti, Cameroon as well as 10 isofemale lines of \Dsim {} from Nairobi, Kenya and 10 isofemale lines from Madagascar.  Duplications were identified through divergently oriented reads and coverage changes in comparison to reference genomes.  We identify 1415 tandem duplications in \Dsim {} and 975 tandem duplications segregating in \Dyak {} that span 845 different gene sequences in \Dyak {} and 478 different gene sequences in \Dsim {} \citep{Rogers2014}.  Gene duplications were defined as any divergent read calls whose maximum span across all lines overlaps with the annotated CDS coordinates. \Dyak {} CDS annotations were based on flybase release \Dyak {} r.1.3.  Gene annotations for the recent reassembly of the \Dsim {} reference were produced by aligning all \Dmel {} CDS sequences to the \Dsim {} reference in a tblastx.  Percent coverage of the CDS was defined based on the portion of the corresponding genomic sequence from start to stop that was covered by the maximum span of divergent read calls across all strains.    Using the representation of gene sequences in \Dyak {} of $\frac{845}{16082}$ we use a binomial test to calculate the likelihood of 56 shared variants among the 478 genes duplicated in \Dsim. 
\subsection*{Frequency of unsampled alleles}
We also estimated the frequency distribution for alleles that are absent among our 20 strains, according to a Bayesian binomial model.  Assuming that sampling follows a binomial model that is dependent upon the allele frequency $p$, the probability that a variant is present at frequency $p$ given that it is not observed is as follows:

 $P(p | absent)= \frac {p(absent | p) *p( p)}{\int_0^{1} p(absent |p)*p ( p) \mathrm{d} p.}$

Assuming a uniform distribution on $p$:

$P(p| absent)= \frac {p(absent | p)}{\int_0^{1} p(absent |p) \mathrm{d} p.}$

$P(p | absent)= \frac {(1-p)^{20}}{\int_0^{1} (1-p)^{20} \mathrm{d} p.}$

$P(p | absent)= 21 (1-p)^{20}$

95\% lower CI  defined as: 

$\int_0^x {21 (1-p)^{20}} \mathrm{d}p=0.95$ 

$1- (1-x)^{21} =0.95$

And therefore the 95\% one-sided lower CI is $x \leq 0.132946 $ whereas the 50\% one-sided lower CI will be $x \leq 0.0325$.  Placing a uniform prior on $p$ will bias estimates toward higher frequency variants, thereby placing a conservative upper bound on allele frequencies.

\subsection*{Estimated number of segregating tandem duplications}

We compared the estimated total number of duplications expected in a population to estimates of diversity based on our sample of 20 strains, correcting $S$ for a 3.9\% false positive rate (Table \ref{SegSites}).  Under a standard coalescent model \citep{Wakeley2009,Ewens1974,Watterson1975}:  

$E[S_{population}]=\frac{S_{sample}}{a_{sample}}*a_{population}$ 

Where a in a sample of size n (in this case n=20):

$a_{sample}=\sum_{i=0}^{n-1} \frac{1}{i}$

$a_{population}=\sum_{i=0}^{2N_e} \frac{1}{i} $

When $2N_e$ is large:

$\sum_{i=0}^{2N_e} \frac{1}{i} \approx \theta  (ln(2N_e) +0.57722)$

Hence:

$E[S_{population}]=\frac{S_{sample}}{a_{20}} *(ln(2N_e) +0.57722)$

We can use similar methods to estimate the variance in the number of segregating sites in the population.

$Var[S_{population}]=\theta \sum_{i=0}^{2N_e} \frac{1}{i}+ \theta^2 \sum_{i=0}^{2N_e} \frac{1}{i^2}$

When $2N_e$ is large:

$ \sum_{i=0}^{2N_e} \frac{1}{i^2}\approx \frac{\pi^2}{6}$.

$Var[S_{population}]=\theta(ln(2N_e) +0.57722)+ \theta^2\frac{\pi^2}{6}$

\subsection*{Gene Ontology}

Overrepresented functional categories were identified using DAVID gene ontology software with an EASE threshold of 1.0. as previously described \citep{Rogers2014}.  We observe several functional categories indicative of rapid evolution that are shared between the two species (Table \ref{GOBiasDual}).  As a comparison, we selected a random subset of 845 {} genes for \Dyak and 478 genes from \Dsim, and performed ontology analysis for a comparison.  In \Dyak, male courtship, GTPase enzymes, and alt splicing had significant group EASE thresholds whereas \Dsim showed marginal values on esterases, tracheal development, neurodevelopment, lipid metabolism, hormone receptors, cell communication and growth and starvation.   Nothing has an EASE of 1.5 in either species similar to the values observed in Table \ref{GOBiasDual}. There is no agreement in functional categories for the two species in the random subsets.  Thus, we would suggest that the convergence across species and the associations with rapid evolution are not the product of sampling errors. 

\subsection*{Proportion of the genome represented by segregating duplicates} 
To determine the number of duplications necessary to span the full range of the genome, we simulated chromosomes with a length determined by the number of base pairs with non-zero coverage in our reference strain.  We then simulated random draws from the distribution of duplication lengths for each chromosome, placing duplication start sites at random and recorded the number of duplications necessary to cover 10\%, 25\%, 50\%, and 90\% of sequence length for each chromosome in each trial.  Simulations were repeated for 1000 trials for each chromosome.  

These simulations do not account for mutational biases that might result in clustering of duplications in particular regions while other regions remain static, nor do they require that new duplications reach an appreciable frequency so that they are immune to stochastic loss through genetic drift.  They do not require that duplications capture sufficient sequence to have functional impacts or require that breakpoints not disrupt known functional elements.  Furthermore, simulating individual chromosomes separately decreases the likelihood of resampling particular sites thereby lowering the estimated number of duplications needed to cover the entire genome.  Hence, these estimates put a highly conservative lower bound on the minimum number of mutations necessary to capture the full genomic sequence.

To estimate the expected proportion of the genome spanned by all duplicates in the population, we resampled 6700 duplicates from the observed size distribution of \Dyak {} with replacement and 4000 duplicates from the observed size distribution of \Dsim, placing duplications at random positions across the chromosome.  We performed 100 replicates of sampling and report the mean across all replicates for each species.     In \Dsim {} we observe one case with 19 independent whole gene duplications of a single ORF \citep{Rogers2014}, suggesting up to 1000-fold variation in mutation rates over the genome average.  Estimates of population level variation and genome wide mutation rates ignore mutation rate variation where some regions may be highly prone to duplications whereas others remain static, which would reduce likelihood of unobserved tandem duplications outside of mutational hotspots.  Hence, these estimates represent a lower bound on the number of duplications necessary to span the entire genome.  

\subsection*{Mutation rates and wait times for duplicates}
We estimate average heterozygosity ($\theta_{\pi}$) and mutation rates ($\mu$) per gene for \Dyak {} and \Dsim {} for gene duplications that capture at least 90\% of gene sequence (in agreement with previous estimates \citep{RBH}), for genes that recruit non-coding sequence, and for chimeric genes.   Heterozygosity estimates used to calculate mutation rates were corrected for ascertainment bias (see SI Text) and excluded genes that were captured by 4 or more independent mutations, a signal of hotspots and mutation rate heterogeneity.  Heterozygosity per gene is estimated given 16,082 gene sequences in \Dyak {} and 10,786 coding sequences in \Dsim {} (Table \ref{MutRates}).   Given estimates of $\theta_{\pi}$, we estimate the probability of adaptation from standing variation under an additive genetic model for neutral variants,  $P_{sgv}=e^{-\theta_{\pi} *ln(1+2N_{e}s)}$ \citep{Hermisson2005} and the time to establishment ($T_e$) of a deterministic sweep from new mutations, such that new mutants escape the stochastic forces of drift, $T_e=\frac{1}{\theta_{\pi}s}$ \citep{Maynard1971,Gillespie1991}.  These estimates are provided for two strong selection coefficients of $s=0.01$ typical of what is observed in \Dros {} and $s=0.20$ modeling abnormally strong selection on a single locus consistent with \cite{Karasov2010}.   Estimates assume that a given site of interest is under strong selection and that the beneficial mutation rate is equal to the mutation rate per site per generation providing an upper limit on the ability of new mutations to facilitate adaptation.  In reality, strongly selected sites will be rare throughout the genome \citep{Jensen2008,Andolfatto2007} and these wait times given strong selection will not reflect the expected number of selective sweeps throughout the genome.

\subsection*{Additional methods}
Further description of methods including description of intronic SNPs, analysis of population structure, residual heterozygosity, calculation of $N_e$, and correction for ascertainment bias are available in SI Text.  All data files are available via http://molpopgen.org/Data and http://www.github.com/ThorntonLab/DrosophilaPopGenData-Rogers2014.  Aligned bam files were deposited in the National Institutes of Health Short Read Archive under accession numbers SRP040290 and SRP029453.  Sequenced stocks were deposited in the University of California, San Diego (UCSD) stock center with stock numbers 14021-0261.38- 14021-0261.51 and 14021-0251.293 - 14021-0251.311.

\section*{Acknowledgements}
The authors would like to thank Nigel F. Delaney, Elizabeth G. King, Anthony D. Long, and Alexis S. Harrison, and Trevor Bedford  for helpful discussions as well as three anonymous reviewers whose comments substantially improved the manuscript.  RLR is supported by NIH Ruth Kirschstein National Research Service Award F32-GM099377.  Research funds were provided by NIH grant R01-GM085183 to KRT and R01-GM083228 to PA.  All sequencing was performed at the UC Irvine High Throughput Genomics facility, which is supported by the National Cancer Institute of the National Institutes of Health under Award Number P30CA062203.  The content is solely the responsibility of the authors and does not necessarily represent the official views of the National Institutes of Health.  The funders had no role in study design, data collection and analysis, decision to publish, or preparation of the manuscript.   RLR, JMC, KRT, and PA performed analyses.  LS generated Illumina sequencing libraries. TTH provided gene annotations for \Dsim.  RLR, JMC, PA and KRT designed experiments and analyses.
}

\bibliographystyle{MBE}
\bibliography{PopGenMain}

\newpage

\clearpage

\begin{table}
\begin{threeparttable}
\begin{center}
\caption{\label{MutRates} Establishment of Sweeps in \Dyak {} and \Dsim. }
\begin{tabular}{llrrrr}
 \hline
Species &  & Intron SNPs &Whole Gene  & Recruit & Chimera \\
\hline
\Dyak&$\mu$ & $5.8\times10^{-9}$& $1.17 \times 10^{-9}$ &  $3.46 \times 10^{-10}$ &  $3.70 \times 10^{-10}$\\
  &$\theta_{\pi}$ & 0.0138 & 0.00277 &  0.00082 & 0.00088 \\
\\
&$P_{sgv}$,  s=0.01 & 12.1\%  & 2.23\%  & 0.67\%  & 0.71\% \\
&$P_{sgv}$, s=0.20 & 15.7\% & 3.05\% & 0.91\% & 0.97\% \\
\\
& $T_e$, s=0.01 & 7270 & 36,000 & 122,000 & 114,000 \\
& $T_e$, s=0.20 & 364 & 1,800 & 6,087 & 5,704 \\
\\
 \hline
Species &  & Intron SNPs &Whole Gene  & Recruit & Chimera \\
\hline
\Dsim & $\mu$ & $5.8\times10^{-9}$ &  $6.03 \times 10^{-10}$ &  $2.42 \times 10^{-10}$ &  $8.52 \times 10^{-11}$\\
& $\theta_{\pi}$ & 0.0280 & 0.00291 &  0.00117 & 0.00041 \\
\\
&$P_{sgv}$, s=0.01 &  24.6\% &   2.56\% & 1.04\%  &  0.37\% \\
&$P_{sgv}$, s=0.20 & 30.1\%  & 3.41\%  & 1.38\% & 0.49\%  \\
\\
& $T_e$, s=0.01 & 3580 & 34,400 & 85,700 & 243,000 \\
& $T_e$, s=0.20 & 179 & 1,720 & 4,290 & 12,100 \\
\hline
\end{tabular}

\begin{tablenotes}
\item[]$P_{sgv}$ from Hermisson and Pennings (2005) estimates the likelihood of adaptation from standing genetic variation under an additive model assuming neutral variation.
\item[] $T_e$ (Gillespie 1991 and Maynard-Smith 1971) estimates the average time until establishment of a selective sweep from a new mutation in generations given that a site is under strong selection with beneficial mutation rate equal to $\theta_{\pi}$.  Estimates provide a lower bound on $T_e$.
\end{tablenotes}
\end{center}

\end{threeparttable}
\end{table}

\clearpage

\begin{figure}[h]
\begin{center}

\includegraphics[scale=0.60]{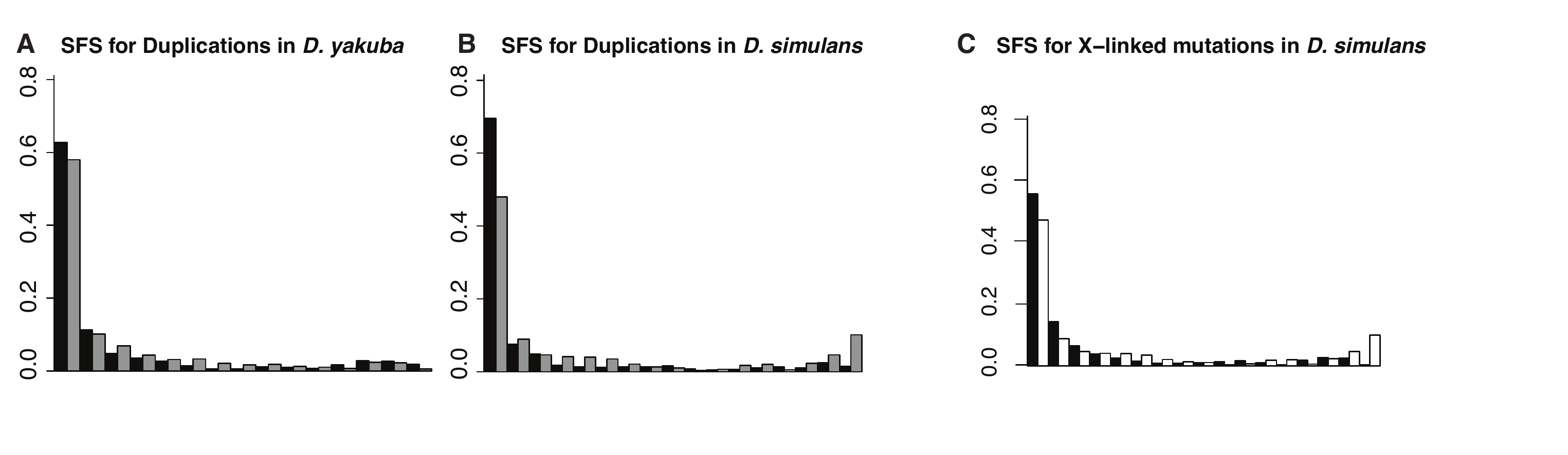}
\caption{\label{SFSSingleSm}SFS for tandem duplications in \Dyak {} and \Dsim, corrected for ascertainment bias. A. Site frequency spectra on the autosomes (black) and on the X (grey) in \Dyak. B.  SFS on the autosomes (black) and on the X (grey) in \Dsim.  C. SFS for X-linked intronic SNPs (black) and duplicates (grey).  The excess of high frequency variants on the X in \Dsim {} suggests widespread selection for tandem duplicates on the \Dsim {} X.  }
\end{center}
\end{figure}
\clearpage

\begin{figure}[h]
\begin{center}

\includegraphics[scale=0.60]{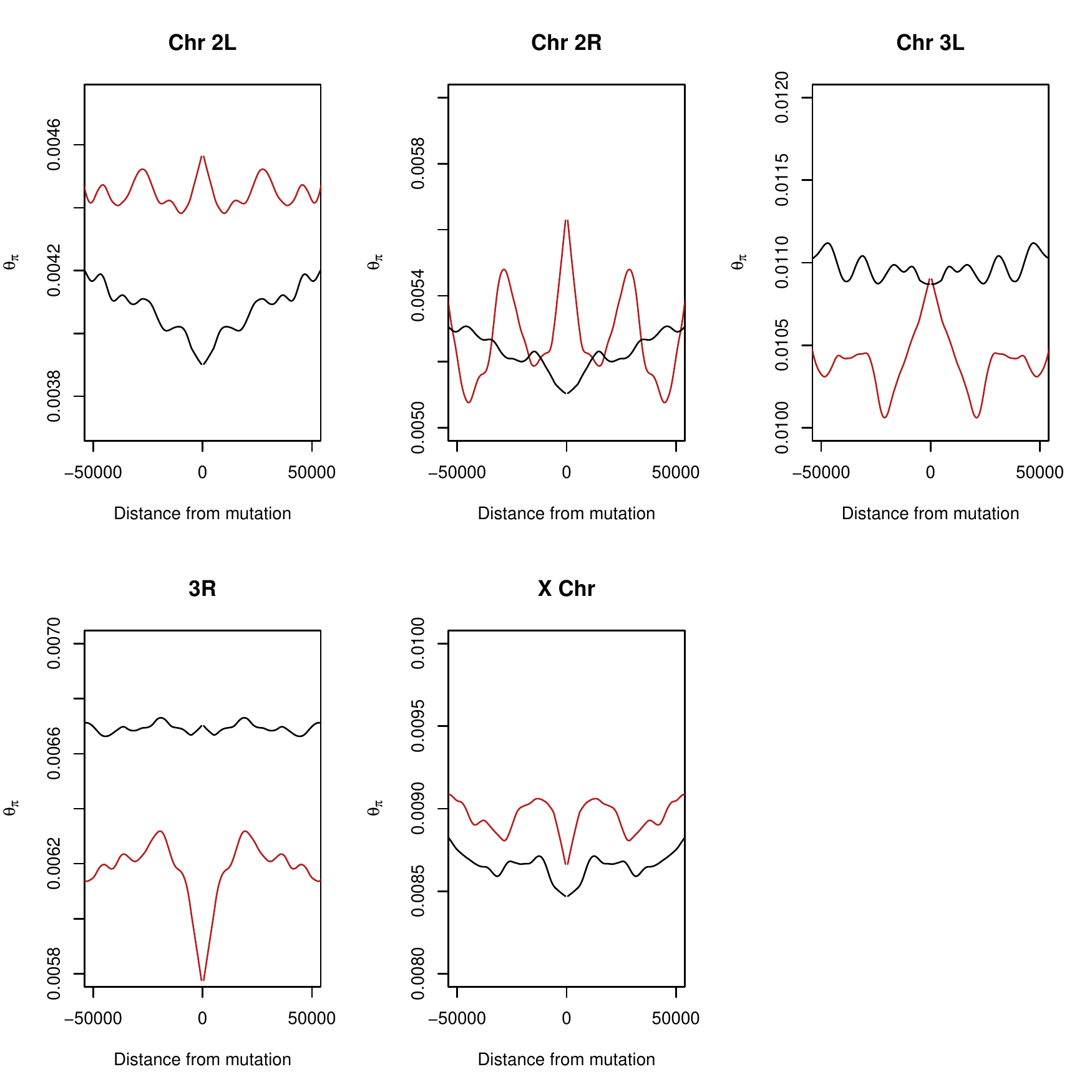}
\caption{\label{DistanceFromDyak}Diversity ($\theta_{\pi}$) as a function of distance from new mutations in \Dyak {} for putatively neutral intronic SNPs (black) and for tandem duplications (red) by chromosome with lowess smoothing.  Duplicates show a reduction in diversity approaching duplications, whereas neutral SNPs show no reduction in diversity.  Plots exclude centromeric regions and the 4th chromosome which have atypical nucleotide diversity.  }
\end{center}
\end{figure}
\clearpage

\begin{figure}[h]
\begin{center}

\includegraphics[scale=0.60]{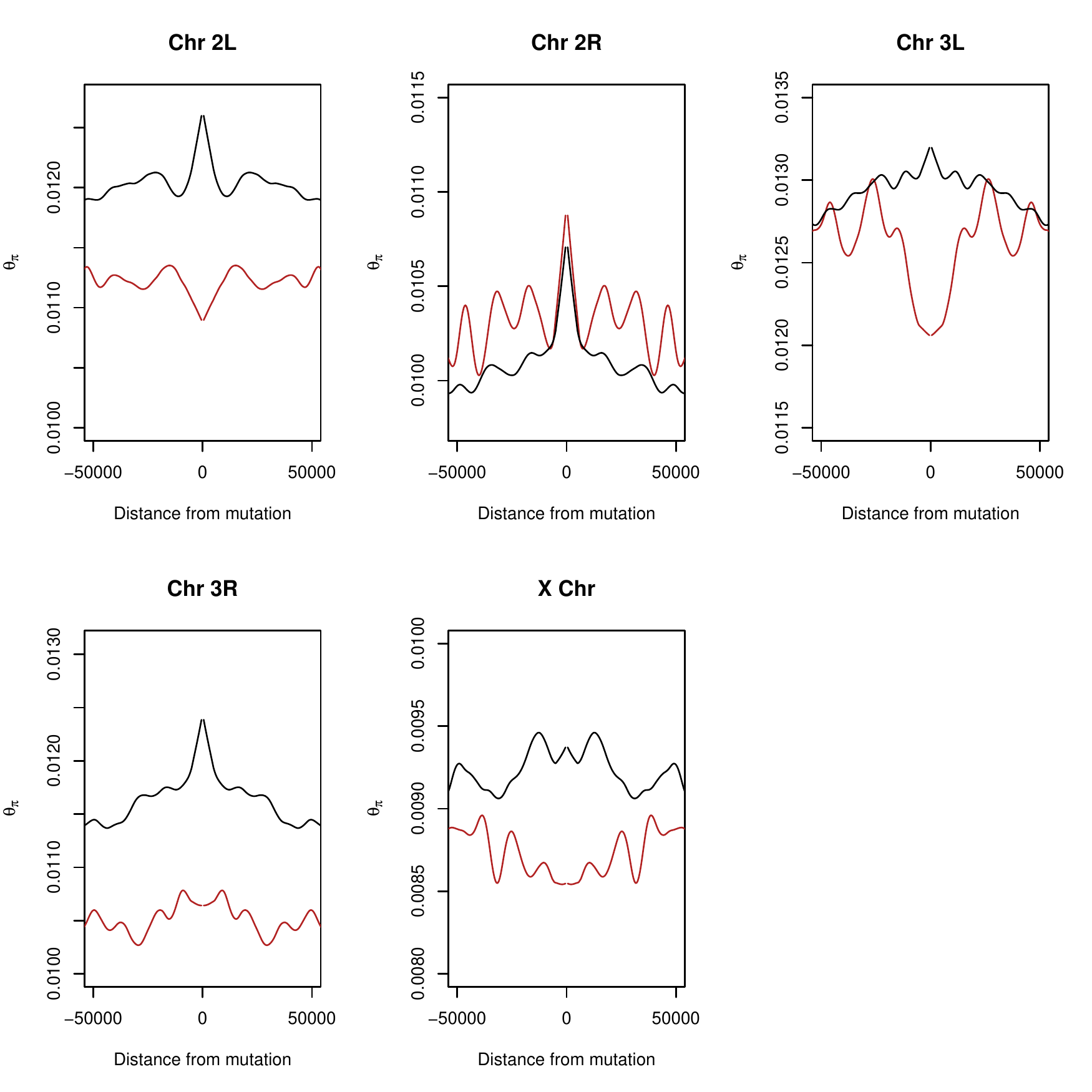}
\caption{\label{DistanceFromDsim} Nucleotide diversity, $\theta_{\pi}$ as a function of distance from new mutations in \Dsim {} for putatively neutral intronic SNPs (black) and for tandem duplications (red) by chromosome with lowess smoothing.  Duplicates show a reduction in mean diversity approaching duplications, whereas neutral SNPs show no reduction in diversity.  Plots exclude centromeric regions and the 4th chromosome which have atypical nucleotide diversity.  Chromosome 3L is strongly affected by a cluster of duplications at roughly 8.5Mb, which is excluded from the plot. }
\end{center}
\end{figure}

\clearpage

\begin{figure}[h]
\begin{center}

\includegraphics[scale=0.80]{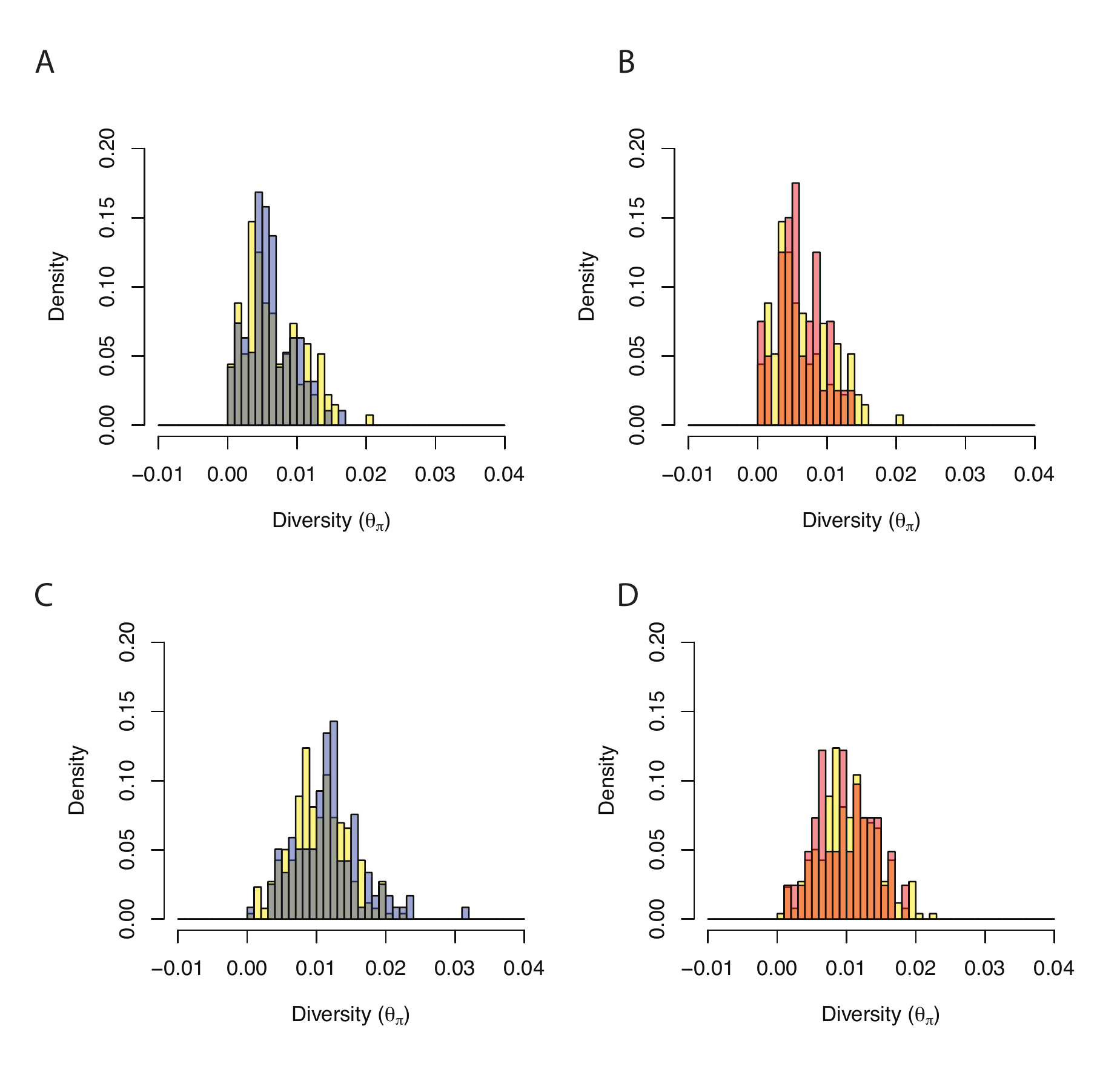}
\caption{\label{ThreeTypes} Histogram of nucleotide diversity, $\theta_{\pi}$, (A) For Intergenic mutations (yellow) and duplications that capture gene sequences but do not create chimeric constructs (blue) in \Dyak.(B) For Intergenic mutations (yellow) and duplications that create chimeric genes (red) in \Dyak. (C) For Intergenic mutations (yellow) and duplications that capture gene sequences but do not create chimeric constructs (blue) in \Dsim.  (D) For Intergenic mutations (yellow) and duplications that create chimeric genes (red) in \Dsim. }
\end{center}
\end{figure}

\clearpage

\begin{figure}
\centering
  \includegraphics[width=\linewidth]{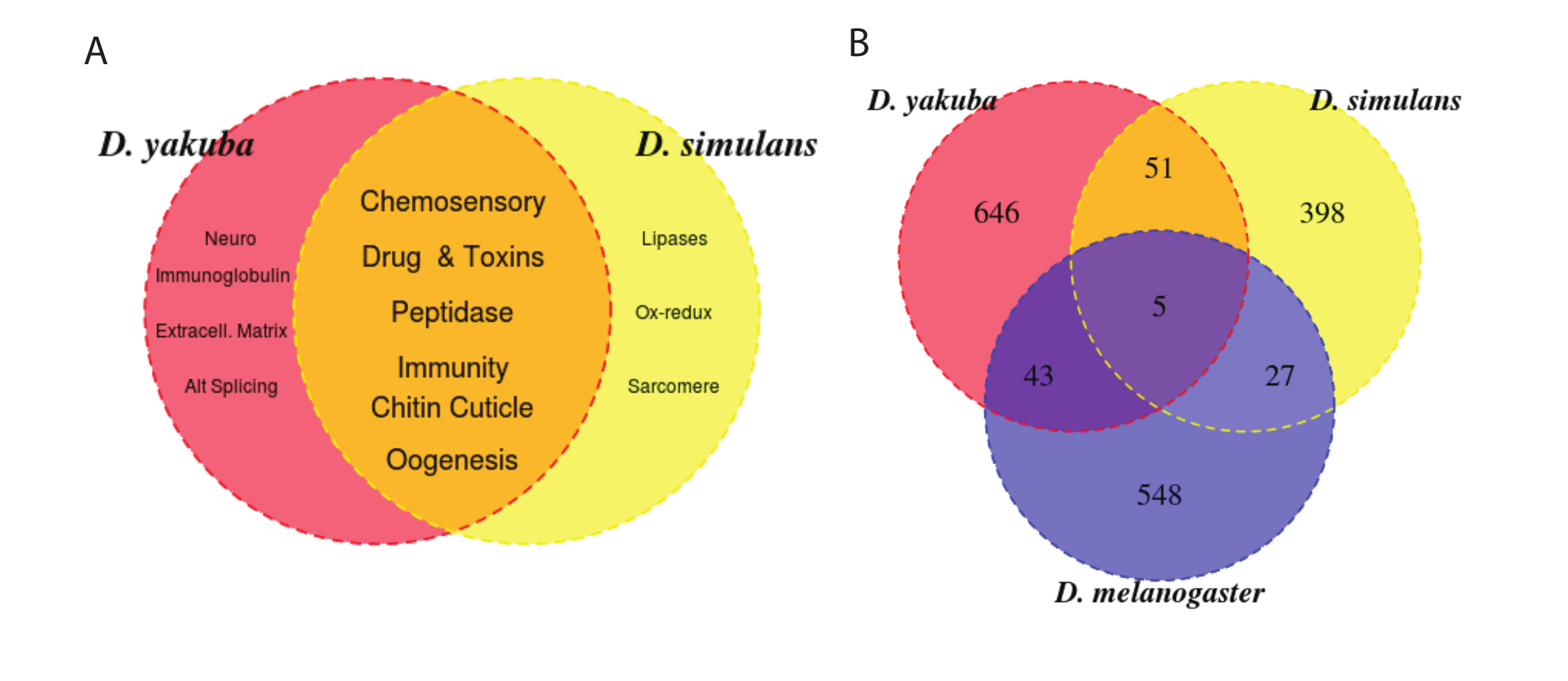}

\caption{\label{VennDual} A) Gene ontology classes overrepresented by species among singly duplicated genes or among multiply duplicated genes.   B) Number of genes duplicated by species. Most variants are species specific, with small numbers of parallel duplication of orthologs across species. }

\end{figure}
\clearpage

\begin{figure}
\includegraphics[scale=.8]{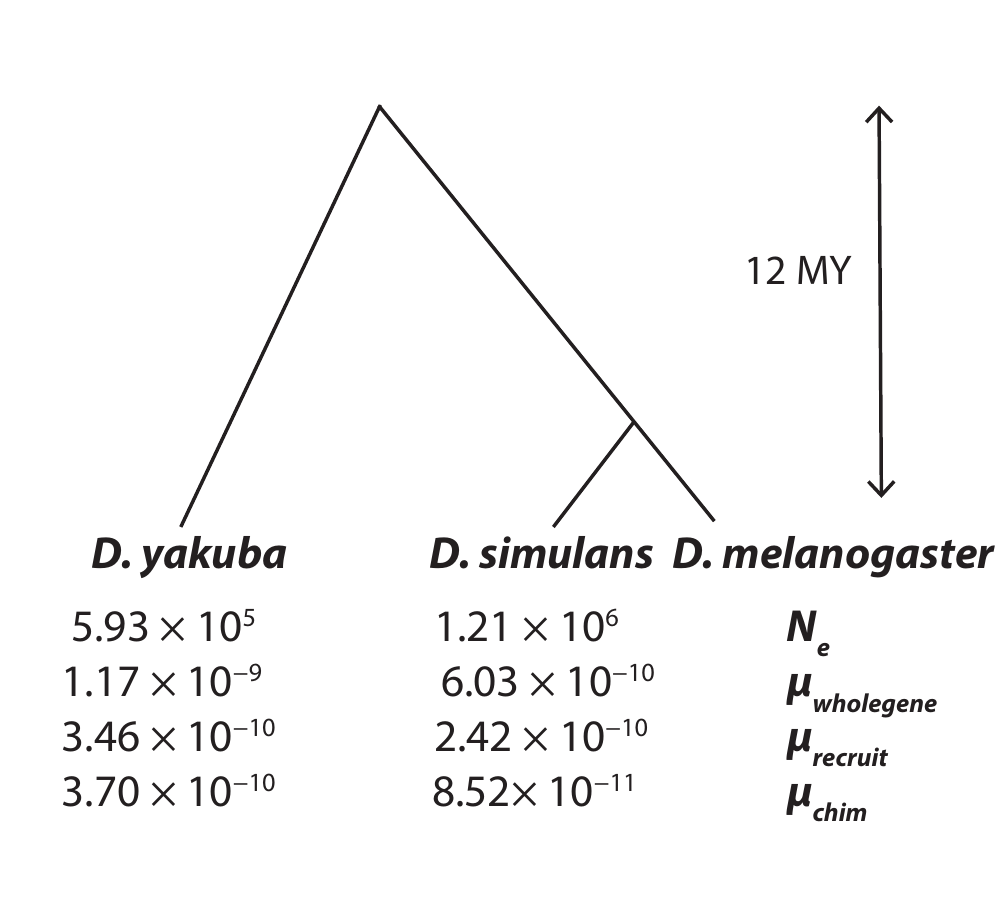}
\caption{\label{PhyloNums} Genomewide population mutation rates for all duplications ($\theta$), population sizes ($N_e$), and per gene mutation rates ($\mu$) for gene structures produced by duplication by species.  Low mutation rates and mutation limited evolution leads to low levels of parallel recruitment of tandem duplications. }
\end{figure}

\clearpage

%
%
%


\chapter*{}
\section*{Supporting Information}
\renewcommand{\thefigure}{S\arabic{figure}}
\renewcommand{\thetable}{S\arabic{table}}
\setcounter{figure}{0}
\setcounter{table}{0}
\setcounter{page}{1}

{\setlength{\baselineskip}
{2.0\baselineskip}
\renewcommand{\baselinestretch}{2.0}

\subsection*{Identifying duplicated coding sequence} 

Tandem duplications were previously identified using a combination of paired end read mapping and coverage changes in 20 isofemale lines of \Dyak {} and 20 isofemale lines in \Dsim {} generated via 9-12 generations of sibling mating from wild-caught flies.  We sequenced 10 isofemale lines of \Dyak {} from Nairobi, Kenya, and 10 isofemale lines from Nguti, Cameroon as well as 10 isofemale lines of \Dsim {} from Nairobi, Kenya and 10 isofemale lines from Madagascar.  Duplications were identified through divergently oriented reads and coverage changes in comparison to reference genomes.  We identify 1415 tandem duplications in \Dsim {} and 975 tandem duplications segregating in \Dyak {} that span 845 different gene sequences in \Dyak {} and 478 different gene sequences in \Dsim {} \citep{Rogers2014}.  Gene duplications were defined as any divergent read calls whose maximum span across all lines overlaps with the annotated CDS coordinates. \Dyak {} CDS annotations were based on flybase release \Dyak {} r.1.3.  Gene annotations for the recent reassembly of the \Dsim {} reference were produced by aligning all \Dmel {} CDS sequences to the \Dsim {} reference in a tblastx.  Percent coverage of the CDS was defined based on the portion of the corresponding genomic sequence from start to stop that was covered by the maximum span of divergent read calls across all strains.    Using the representation of gene sequences in \Dyak {} of $\frac{845}{16082}$ we use a binomial test to calculate the likelihood of 56 shared variants among the 478 genes duplicated in \Dsim. 

\subsection*{Intronic SNPs}
In order to produce a neutral proxy for sequence change in each species, we identified SNPs for short introns 100 bp or less, focusing on sites 8-30 which are generally subject to little constraints \citep{Halligan2006, Parsch2010, Clemente2012}.  Reads containing indels were re-aligned using GATK \citep{GATK}.  SNPs were identified across strains using samtools v1.18  mpileup \citep{samtools} disabling probabilistic realignment (-B) and outputting genotype likelihoods in BCF format (-g).  The resulting BCF used to create a VCFusing bcftools, calling bases using Bayesian inference (-c) calling genotypes per sample (-g) with a scaled mutation rate of 1\% (-t .01) under a haploid model (ploidy=1).   SNPs were required to have minimum Illumina coverage depth of 20 reads, maximum coverage of 250 reads,Ê{} MQ$\geq20$, and GQ$\geq30$Ê{} andÊ{} invar GQ$\geq40$. We excluded SNPs identified in the reference, which are indicative of either assembly errors or residual heterozygosity.  We performed hierarchical cluster analysis in R using all SNPs by chromosome to evaluate population structure.  

The ancestral state for each SNP was established through comparison with the nearest sequenced reference genome as an outgroup, \Dere {} for \Dyak {} sequences and \Dmel {} for \Dsim {} sequences.  Orthologs between each species and its outgroup were identified using reciprocal best hit criteria in a BLASTn at an E-value cutoff of $10^{-5}$.  Full gene sequences for each ortholog were then aligned using clustalw, keeping only genes which aligned with 85\% or greater nucleotide identity.  Divergence between the two species, $Div_{x,y}$, was defined based on alignments of intronic sites from bases 8-30 between each species and the outgroup reference genome, excluding gapped sequences, for aligned orthologs with 85\% or nucleotide identity. The ancestral state was defined based on the corresponding sequence in the outgroup genome.  We excluded sites where the outgroup reference was in disagreement with both the \Dyak {} reference and \Dyak {} SNPs, as well as triallelic SNPs, sites with reference sequence of `N', or SNPs identified in the VCF for the reference, suggesting inaccuracies in reference assembly or residual heterozygosity in the reference.  These resulted in a total of 7158 intronic SNPs in \Dyak {} and 5504 intronic SNPs in \Dsim. The resulting unfolded SFS was then corrected for the probability of independent mutations in both reference genomes leading to incorrect inference of the ancestral state. 

Given net divergence $D_{net} = Div_{x,y} - \pi_{x}$, the probability of identical independent mutations occurring in the outgroup reference genome is reflected by either the probability of an independent transition (ts) at the site of a transition mutation, or by 1/2 the probability of a transversion (tv) at the site of a transversion polymorphism.  Thus, 

\begin{equation}\label{keq} k=  [(\frac{\kappa}{2+\kappa})^2 +\frac{1}{2}(\frac{2}{2+\kappa})^2] D_{net}  \end{equation}

Empirically, in \Dros {} $\kappa =\frac{ts}{tv}=2$.  Thus, $k=\frac{3}{8}D_{net}$.

 The unfolded SFS for intronic sites was corrected for the likelihood of independent mutations in the reference, k.  The probability of independent mutations occurring in both genomes is equal to the probability of either two independent transitions or two independent transversions occurring in both genomes. We calculated $\pi_x$ as the average heterozygosity per intronic site.

  Given a likelihood of independent identical mutations of $k=\frac{3}{8} D_{net}$ (See Text S1).
\begin{equation}\label{eqA} S_{i,obs}= E[S_i] - E[S_i](k) +E[S_{n-i}](k) \end{equation}
\begin{equation}\label{eqB} S_{n-i,obs}=E[S_{n-i}] - E[S_{n-i}](k) +E[S-i](k) \end{equation} 

Substituting Equation \ref{eqB} into Equation \ref{eqA}, we obtain \begin{equation} E[S_i] = \frac{S_{i,obs}(1-k) - S_{n-i,obs}(k)}{1-2k} \end{equation}

\subsection*{Correcting Duplicates for Ascertainment Bias}
Tandem duplications, unlike SNPs, cannot be identified using paired-end reads in individual strains except through comparison to the reference genome.  Moreover, variants that are segregating at high frequency in populations are substantially more likely to be present in the reference, and therefore are substantially less likely to be identified in sample strains \citep{Emerson2008}.  We corrected site frequency spectra according to the model developed by Emerson et al. \citep{Emerson2008}.

\begin{equation}  x_i= \frac{y_i  \frac{n}{n-i} }{\sum_{i=1}^{n-2}{y_i  \frac{n}{n-i}  }  } \end{equation}

Here, $x_i$ is the true proportion of alleles at frequency i in the population, and $y_i$ is the observed proportion of alleles at frequency i in a sample of n strains (here 21).  The correction for ascertainment bias lowers estimates of the proportion found a low frequencies and increases estimates of the proportion at high frequency.  For estimates of population site frequency spectra, we removed all variants with divergently oriented reads in the reference strain, as these would not be identified in an accurately annotated reference.

\subsection*{Residual heterozygosity}
Some isofemale lines contained regions of residual heterozygosity in spite of over 10 generations of inbreeding in the lab.  To detect regions of residual heterozygosity, we called SNPs as above under a diploid model.  Segments with residual heterozygosity were detected using an HMM (ÒHMMÓ;http://cran.r-project.org/web/packages/HMM/).  
\\
Prior probabilities on states were set as:

$\pi=\begin{bmatrix}
0.5.  & 0.5\\
\end{bmatrix}$
\\
Transition probabilities were set to: 

$T = \begin{bmatrix}
1-10^{-10}&10^{-10} \\
10^{-10} &1-10^{-10} \\
\end{bmatrix}$\\

and emission probabilities set to:

$E = \begin{bmatrix}
\theta&\epsilon \\
1-\theta&1-\epsilon\\
\end{bmatrix}$

Where $\epsilon=0.001$  and  $\theta=0.01$. The most likely path was calculated using the Viterbi algorithm, and heterozygous segments 10kb or larger were retained. Heterozygous blocks within 100kb of one another in a sample strain were clustered together as a single segment to define the span of residual heterozygosity within inbred lines.   

 \subsection*{Differences in Site Frequency Spectra}

If different classes of duplications have different selective impacts, we should observe clear differences in site frequency spectra, with more positively selected duplications showing fewer singleton alleles and more high frequency variants.  Site frequency spectra are not normally distributed, nor can they be normalized through standard transformations, and thus require non-parametric tests. We used a two-sided Wilcoxon rank sum test to determine whether site frequency spectra were significantly different.    For each comparison, we excluded tandem duplications that are present in the reference genomes as well as putative ancestral duplications, as these are likely to display biases with respect to size, propensity to capture coding sequences, and association with repetitive content. We compared site frequency spectra of the following groups within each species: duplications on the X and on the autosomes and all pairwise combinations of SNPs and duplicates on the X and autosomes.  We also performed Kolmogorov-Smirnov test for comparison.  In \Dsim, we used a $\chi^2$ test to determine whether high frequency alleles are overrepresented among duplications on the X relative to intronic SNPs, comparing the proportion of variants as at a sample frequency of $\geq \frac{16}{17}$.  
 
 Tandem duplicates that lie in regions with residually heterozygous segments extending 1kb upstream or downstream were excluded from the SFS, resulting in unequal sample sizes for different variants.  Samples with fewer than 15 strains remaining were excluded from the SFS.  The SFS for intronic SNPs and for duplicates was then scaled to a sample of size 17 in \Dsim {} and 15 in \Dyak {} according to Nielsen et al. \citep{Nielsen2005}.

\subsection*{Segregating Inversions}
In order to check for population substructure, we aligned all SNPs in Intronic sequences from 8-30 bp which are supposed to be a neutral proxy \citep{Halligan2006, Parsch2010, Clemente2012} and performed hierarchical clustering in R using hclust.  These SNPs were intended solely to differentiate strains and were not polarized with respect to the ancestral state or otherwise filtered.  We observe little evidence for population structure in \Dsim {} (Figure \ref{StructureSim}).  However, we identify structure on chromosome 2 in \Dyak {} (Figure \ref{StructureYak}), consistent with known polymorphic inversions prohibiting recombination on chromosome 2 \citep{Lemeunier1976, Llopart2005}.  Strains do not strictly cluster with respect to geography but rather are reticulated amongst other groups.  Moreover, among duplicates we do not observe an excess of moderate frequency alleles as one would expect under population substructure given our sampling scheme (Figure \ref{SFSSNPComp}). Thus, these strains constitute a single admixed population.  

Some strains retained residual heterozygosity even after 9 generations of inbreeding, with greater residual heterozygosity in \Dyak {} than in \Dsim, consistent with inversions segregating in \Dyak.  These regions of residual heterozygosity can result in incorrect estimates of SFS by artificially increasing chances of observing variation.  Site frequency spectra were calculated across all strais by correcting sample frequencies for ascertainment bias, excluding regions of residual heterozygosity and then projecting frequencies onto a sample size of 15 in \Dyak {} and 17 in \Dsim {} according to \citep{Nielsen2005}.   As a neutral  comparison we calculated SFS for intronic SNPs (as above) and projected the SFS down to a sample size of 15 in \Dyak {} and 17 in \Dsim {} (Figure \ref{SFSSNPComp}). 

\subsection*{Likelihood of shared variation through ancestry}
The likelihood of shared variation through shared ancestry can be obtained through a coalescent approach.  The probability that an allele does not coalesce in the time period from the present back to the speciation event that separated \Dyak {} and \Dsim {} is $(1-\frac{1}{2N_e})^{t}$. This can be approximated using $e^{\frac{-t}{2N_e}}$.   We estimate $\theta_{pi}$ for putatively neutral 8-30 bp from short introns using libsequence \citep{libsequence}, ignoring sites that are heterozygous and sites with missing data.  For neutral intronic SNPs, $\theta_{pi}=0.0138$ in \Dyak {} and $\theta_{pi}=0.0280$ in \Dsim.  Using the mutation rate of $5.8\times10^{-9}$ \citep {Haag2007}, we find $ N_e = (0.0138)/(4\times 5.8\times10^{-9})= 5.93\times10^5$ in \Dyak {} and $N_e = (0.0280)/(4\times 5.8\times10^{-9})=1.21\times10^6$ in \Dsim.  Using t=12MY \citep{Tamura2004} and 12 generations per year, and $N_e=1.2\times10^6$ from \Dsim, we obtain a probability of shared ancestry for an allele of $9\times10^{-27}$, vanishingly small.   We have polarized all mutations against the putative ancestral state using outgroup reference genomes, focusing solely on derived mutations \citep{Rogers2014}.  Furthermore, the expectation of shared variation for any two alleles through shared ancestry for \Dyak {} and \Dsim {} is expected to be low.  Even large samples are not expected to harbor shared variation over such timescales \citep{Rosenberg2003}.   Thus, we expect shared variants described here to result from independent mutations, not from long standing neutral polymorphism.  

\subsection*{Estimates of population genetic parameters by chromosome}
We estimate $\theta_{\pi}$, $\theta_{W}$, and \D {} for all SNPs in the \Dyak {} and \Dsim {} genomes, removing sites with missing or ambiguous data as well as heterozygous sites using libsequence \citep{libsequence}.  We calculate $\theta_{\pi}$, $\theta_{W}$, and \D {} for 5 kb windows moving in a 500 bp slide across the genome.  For each window, we divide estimates by the number of sites per window with a minimum Illumina coverage depth of 20 reads, maximum coverage of 250 reads,  MQ$\geq20$, consistent with the threshold used to identify SNPs, in order to estimate $\theta_{\pi}$ and $\theta_{W}$ per site.   We compare $\theta_{\pi}$ per site for regions surrounding derived, segregating tandem duplications with regions surrounding derived, segregating, putatively neutral intronic SNPs from 8-30 bp of short introns, excluding windows with less than 4000 bp out 5000 bp that could be assayed for SNPs.  We exclude second SNPs in a single 5 kb window, and exclude SNPs and duplicates that are found in the centromeric regions, which have unusually low diversity (Figure \ref{ChromPlotsDyak2L}-\ref{ChromPlotsDsimX}).    We compared diversity at the locus of SNPs to diversity at the locus of duplication by chromosome with a Wilcox rank sum test (Table \ref{ReducedDiversityTab}) and estimated mean diversity from 50 kb upstream to 50kb downstream of a mutant (Figure \ref{DistanceFromDyak}-\ref{DistanceFromDsim}).  Tests of nucleotide diversity exclude a cluster of multiple duplications from 8.45 Mb-8.55 Mb which has abnormally low diversity (Figure \ref{ChromPlotsDsim3L}).

}

\clearpage
\bibliographystyle{MBE}
\bibliography{PopGenSupp}
\newpage

\captionsetup[table]{list=yes}

\begin{table}
\caption{Number of duplicated regions detected in \Dyak {} and \Dsim}
\begin{center}
\begin{tabular}{lrr}
& \Dyak & \Dsim \\
\hline
Whole gene& 248 & 296 \\
Partial gene & 745  & 462 \\
Intergenic & 745 & 577  \\
\hline
\end{tabular}
\end{center}
\label{GeneDups}
\end{table} 

\clearpage
\begin{center}
\begin{threeparttable}
\caption{\label{SFSKruskaltab} Wilcoxon Rank Sum Tests of Site Frequency Spectra}

\small
\begin{tabular}{|lllr|c|}
\hline
Species&  Type & Type & $W$ & Adjusted $P$-value \\
\hline
\Dyak& Autosomal SNPs & Autosomal Duplicates & 212 & $3.507\times10^{-6}$**  \\
& X-linked SNPs & X-linked Duplicates & 211 &  $4.781\times10^{-4}$**   \\
&Autosomal Duplicates& X-linked Duplicates & 172 & 0.0128* \\
\hline
\Dsim & Autosomal SNPs & Autosomal Duplicates & 268 & $2.981\times10^{-6}$**  \\
& X-linked SNPs & X-linked Duplicates & 113 &  0.2897 \\
&Autosomal Duplicates& X-linked Duplicates & 183.5 &  0.1848 \\
\hline
\end{tabular}
\begin{tablenotes}
\item[]* $P< 0.05$, ** $P<0.01$
\item[] SNPs are derived from 8-30 bp of short first introns $\leq$ 100bp.
\end{tablenotes}

\end{threeparttable}

 \end{center}
\clearpage
\begin{center}
\begin{threeparttable}
\caption{\label{SFSKolmogorovtab} Kolmogorov-Smirnov Tests of Site Frequency Spectra}

\small
\begin{tabular}{|lllr|c|}
\hline
Species&  Type & Type &  $D$ &Adjusted $P$-value \\
\hline
\Dyak& Autosomal SNPs & Autosomal Duplicates & 0.9333 & $3.868\times10^{-7}$** \\
& X-linked SNPs & X-linked Duplicates & 0.800 & $5.235\times10^{-5}$** \\
&Autosomal Duplicates& X-linked Duplicates & 0.5333 & 0.02625* \\
\hline
\Dsim & Autosomal SNPs & Autosomal Duplicates  & 0.8824 & $4.808\times10^{-7}$** \\
& X-linked SNPs & X-linked Duplicates &0.2941 & 0.4654  \\
&Autosomal Duplicates& X-linked Duplicates & 0.3529 & 0.2402 \\
\hline
\end{tabular}
\begin{tablenotes}
\item[]* $P< 0.05$, ** $P<0.01$
\item[] SNPs are derived from 8-30 bp of short first introns $\leq$ 100bp.
\end{tablenotes}

\end{threeparttable}

 \end{center}
\clearpage
\begin{table}

\caption{\label{ReducedDiversityTab}Reduced Diversity Surrounding Tandem Duplications}
\begin{center}
\begin{tabular}{lrrc}
Species & Chrom & $W$ & single tailed $P$ \\
\hline
\Dyak &  2L & 27768 &  0.9993 \\
& 2R & 27768 & 0.9888 \\
& 3L &  17354.5 &  0.4797 \\
&3R & 76908.5 & 0.0064 \\
& X & 16660 &  0.2674\\
\hline
\Dsim & 2L & 31594 & $1.259\times10^{-4}$\\
& 2R & 56665 &  0.453  \\
& 3L & 41054.5 & 0.01074 \\
& 3R & 42493 & $3.166\times10^{-6}$ \\
& X & 14603.5 & 0.01815 \\
\hline

\end{tabular}
\end{center}
\end{table}
\clearpage

\begin{table}
\caption{ Gene ontology categories overrepresented in both \Dyak {} and \Dsim}
\begin{center}
\begin{tabular}{lrr}
Functional Category & \Dyak {} EASE & \Dsim {}  EASE \\
\hline
Chitin metabolism or cuticle & 2.00 & 0.97 \\
Immune response &  1.44 & 1.59 \\
Drug and toxin metabolism & 1.37 & 2.32 \\
Chemosensation  & 1.12 & 1.37 \\
\hline
Multiple Independent Duplications & & \\
Chorion and oogenesis  & 1.79 & 1.84 \\
Sensory processing &1.23 & 1.41 \\
Immune response & 1.11 & 3.35 \\
\hline
High Frequency Duplicates & & \\
Endopeptidases & - & - \\
\hline
\end{tabular}
\end{center}
\label{GOBiasDual}
\end{table}

\begin{table}
\begin{center}
\caption{\label{SegSites} Estimated Number of Segregating Duplications on X and major Autosomes }
\begin{tabular}{lrr}
 \hline
 &\Dyak &\Dsim \\
 \hline
Genome wide $\theta_W$ for tandem duplications & 383 & 264 \\
$E[S]$ & 5700 & 3800\\ 
$\sigma_S$ & 497 & 344  \\
$E[S] +2\sigma_S$ & 6800 & 4500 \\
Genomic Coverage & 13.4\% & 9.7\%  \\
\hline
\end{tabular}
\end{center}
\end{table}

\clearpage

%
%

\clearpage
\begin{table}
\begin{center}
\caption{\label{Tiled} Number of duplications necessary to cover segments of the genome }
\begin{tabular}{lrr}
 \hline
Percent Covered & Lower Bound (95\% CI) & Upper Bound (95\%CI) \\ 
 \hline
 5\% & 2,358 &2,660 \\
10\% &4,912 & 5,360 \\
25\% &13,668&14,410 \\
50\% & 33,191 & 34,427\\
90\% & 119,799 & 113,767 \\
\hline
\end{tabular}
\end{center}
\end{table}

\clearpage

\begin{table}
\caption{Multiply duplicated genes ($\geq$ 4 duplications) }
\begin{center}
\begin{tabular}{lrr}
Species & Gene & Number \\
\hline
\Dyak & GE10684-PA & 4 \\
&GE13282-PA & 4\\
&GE18810-PA & 4\\
&GE18813-PA & 4\\
&GE18814-PA & 4\\
&GE20773-PA & 4\\
&GE20774-PA & 4\\
&GE25839-PA & 4\\
&GE18811-PA & 6\\
&GE18812-PA & 6\\ 
\hline
\Dsim &CG2174-PD & 4\\
&CG33466-PA & 4 \\
&CG4250-PA & 4 \\
&CG42566-PA & 4 \\
& CG33162-PA &  15 \\
&CG32022-PA &  16 \\
&CG5939-PA &  27 \\
&CG6533-PA &  29 \\
&CG6511-PA &  32 \\
&CG6517-PA &  32 \\
&CG6519-PA &  32 \\
&CG6524-PA &  32 \\
\hline
\end{tabular}
\end{center}
\label{MultipleDups}
\end{table}

\newpage
\section*{Supplementary Figures}
\begin{figure}[h]
\begin{center}

\includegraphics[scale=0.80]{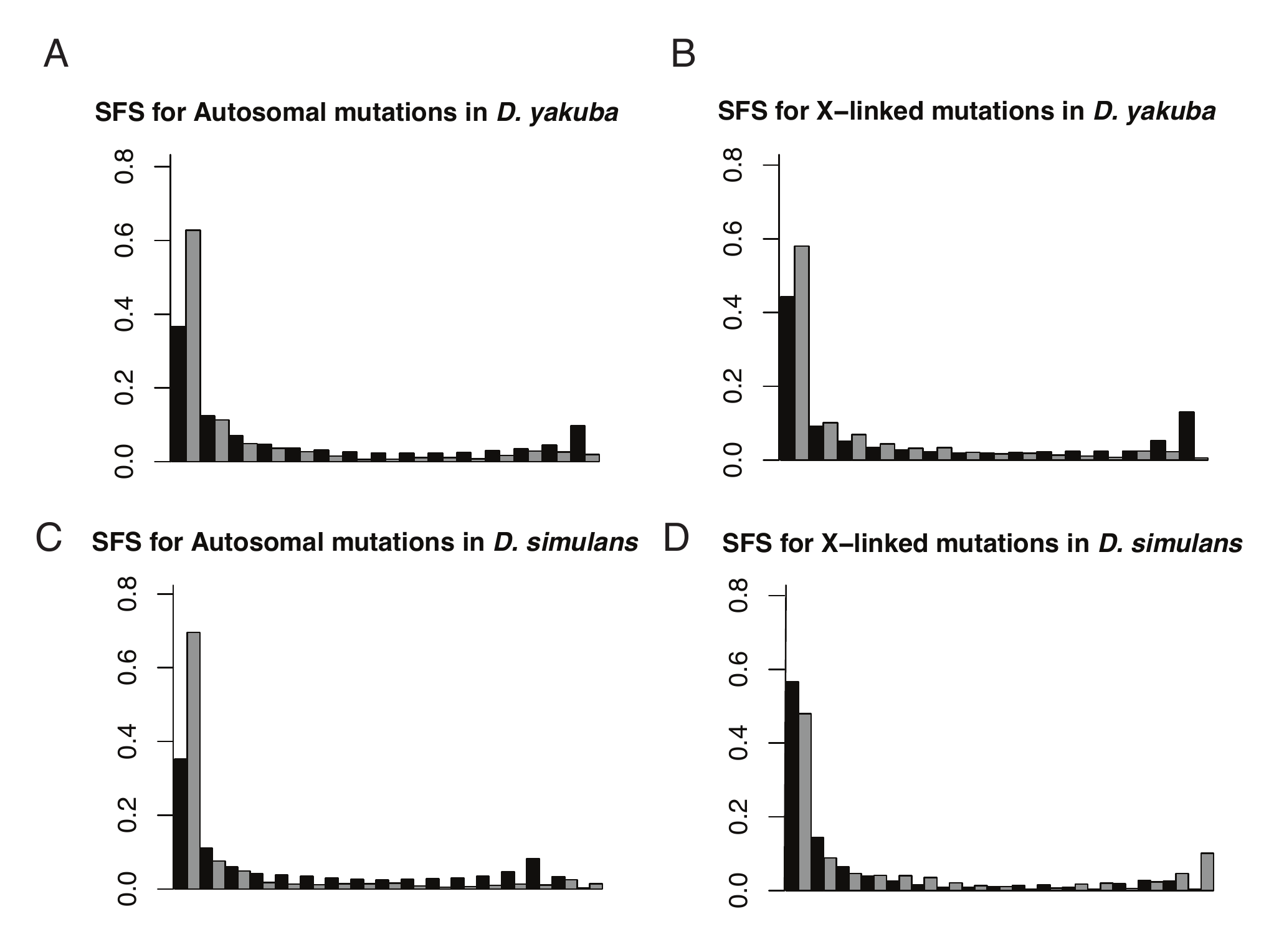}
\caption{\label{SFSSNPComp}Site frequency spectra for SNPs (black) and tandem duplications (grey) on the A) X and B) autosomes in \Dyak {} and on the C) X and D) autosomes in \Dsim.  Tandem duplications in \Dyak {} and on the \Dsim {} autosomes show an excess of low frequency variants, consistent with detrimental phenotypic effects. The \Dsim {} X shows an excess of high frequency variants, consistent with widespread selection favoring duplicates on the X. }
\end{center}

\end{figure}


\clearpage
\begin{figure}[h]
\begin{center}
\includegraphics[scale=1.0]{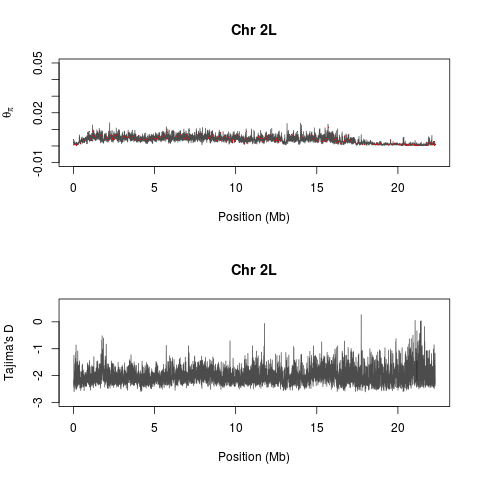}
\caption{\label{ChromPlotsDyak2L} Nucleotide Diversity $\theta_{\pi}$ and \D {} chromosome 2L in \Dyak {} for 5kb windows with a 500 bp slide, showing only windows with 1 kb or more of sequence with coverage sufficient to call SNPS.  \D {} is negatively skewed, consistent with recent population expansion and mean diversity $\theta_{\pi}$ is low surrounding centromeric regions.  Locations and diversity values for duplications are marked in red.}
\end{center}
\end{figure}

\begin{figure}[h]
\begin{center}
\includegraphics[scale=1.0]{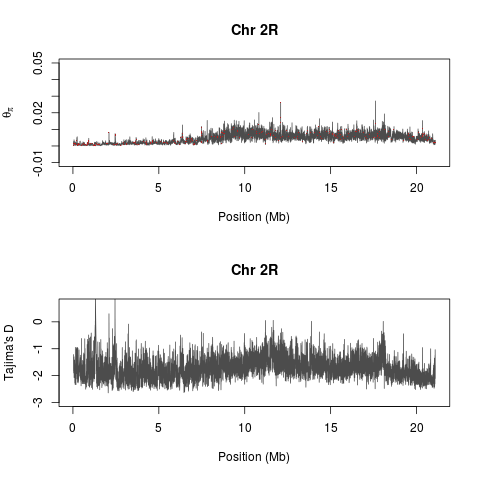}
 \caption{\label{ChromPlotsDyak2R} Nucleotide Diversity $\theta_{\pi}$ and \D {} chromosome 2R in \Dyak {} for 5kb windows with a 500 bp slide, showing only windows with 1 kb or more of sequence with coverage sufficient to call SNPS.  \D {} is negatively skewed, consistent with recent population expansion and mean diversity $\theta_{\pi}$ is low surrounding centromeric regions.  Locations and diversity values for duplications are marked in red.}

\end{center}
\end{figure}

\begin{figure}[h]
\begin{center}
\includegraphics[scale=1.0]{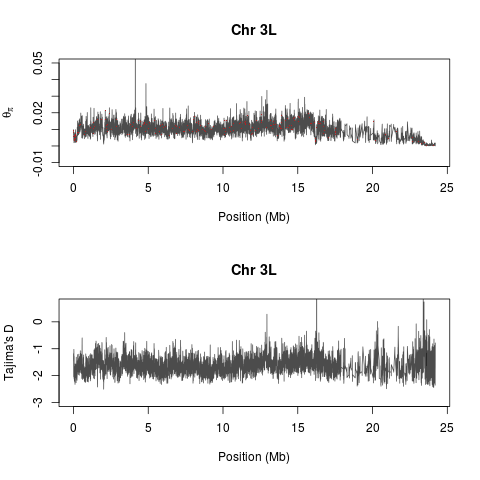}
\caption{\label{ChromPlotsDyak3L} Nucleotide Diversity $\theta_{\pi}$ and \D {} chromosome 3L in \Dyak {} for 5kb windows with a 500 bp slide, showing only windows with 1 kb or more of sequence with coverage sufficient to call SNPS.  \D {} is negatively skewed, consistent with recent population expansion and mean diversity $\theta_{\pi}$ is low surrounding centromeric regions.  Locations and diversity values for duplications are marked in red.}

\end{center}
\end{figure}

\begin{figure}[h]
\begin{center}
\includegraphics[scale=1.0]{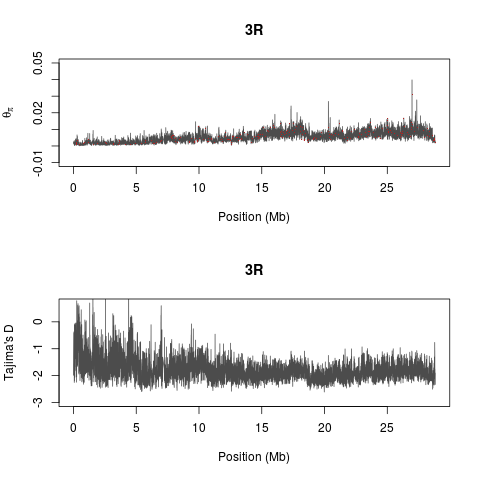}
\caption{\label{ChromPlotsDyak3R} Nucleotide Diversity $\theta_{\pi}$ and \D {} chromosome 3R in \Dyak {} for 5kb windows with a 500 bp slide, showing only windows with 1 kb or more of sequence with coverage sufficient to call SNPS.  \D {} is negatively skewed, consistent with recent population expansion and mean diversity $\theta_{\pi}$ is low surrounding centromeric regions.  Locations and diversity values for duplications are marked in red.}

\end{center}
\end{figure}

\begin{figure}[h]
\begin{center}
\includegraphics[scale=1.0]{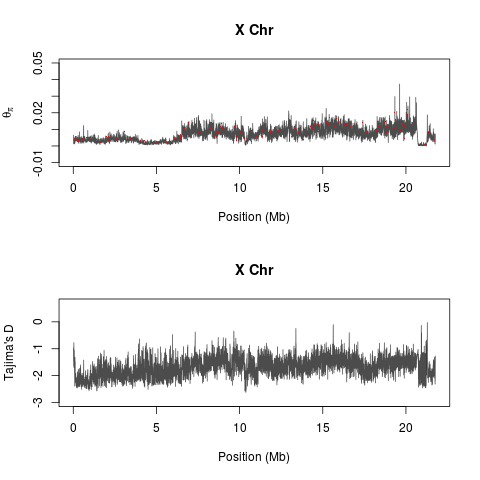}
\caption{\label{ChromPlotsDyak} Nucleotide Diversity $\theta_{\pi}$ and \D {} for the X chromosome in \Dyak {} for 5kb windows with a 500 bp slide, showing only windows with 1 kb or more of sequence with coverage sufficient to call SNPS.  \D {} is negatively skewed, consistent with recent population expansion and mean diversity $\theta_{\pi}$ is low surrounding centromeric regions.  Locations and diversity values for duplications are marked in red.}
\end{center}
\end{figure}

\begin{figure}[h]
\begin{center}
\includegraphics[scale=1.0]{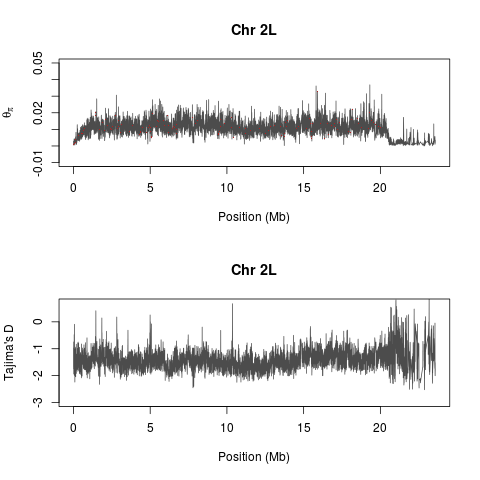}\end{center}
\caption{\label{ChromPlotsDsim2L} Nucleotide Diversity $\theta_{\pi}$ and \D {} chromosome 2L in \Dsim {} for 5kb windows with a 500 bp slide, showing only windows with 1 kb or more of sequence with coverage sufficient to call SNPS.  \D {} is negatively skewed, consistent with recent population expansion and mean diversity $\theta_{\pi}$ is low surrounding centromeric regions.  Locations and diversity values for duplications are marked in red.}

\end{figure}

\begin{figure}[h]
\begin{center}
\includegraphics[scale=1.0]{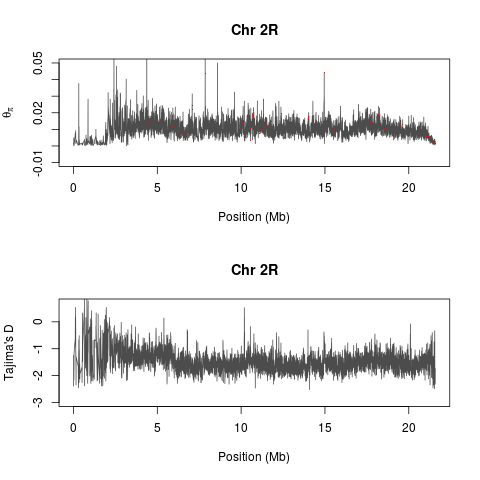}\end{center}
\caption{\label{ChromPlotsDsim2R} Nucleotide Diversity $\theta_{\pi}$ and \D {} chromosome 2R in \Dsim {} for 5kb windows with a 500 bp slide, showing only windows with 1 kb or more of sequence with coverage sufficient to call SNPS.  \D {} is negatively skewed, consistent with recent population expansion and mean diversity $\theta_{\pi}$ is low surrounding centromeric regions.  Locations and diversity values for duplications are marked in red.}

\end{figure}

\begin{figure}[h]
\begin{center}
\includegraphics[scale=1.0]{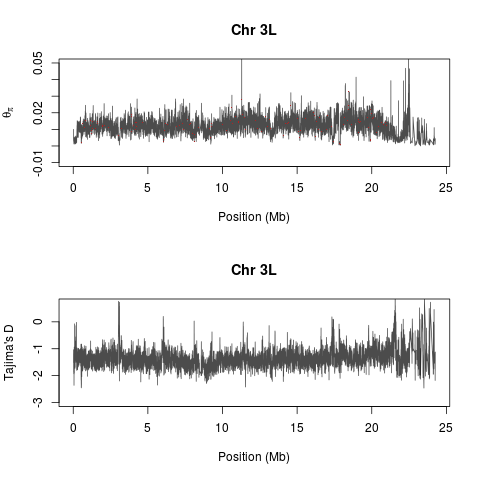}\end{center}
\caption{\label{ChromPlotsDsim3L} Nucleotide Diversity $\theta_{\pi}$ and \D {} chromosome 3L in \Dsim {} for 5kb windows with a 500 bp slide, showing only windows with 1 kb or more of sequence with coverage sufficient to call SNPS.  \D {} is negatively skewed, consistent with recent population expansion and mean diversity $\theta_{\pi}$ is low surrounding centromeric regions.  Locations and diversity values for duplications are marked in red.  Chromosome 3L is strongly affected by a cluster of duplications over 30 independent duplications of a genomic segment at roughly 8.5Mb which shows greatly reduced diversity. }

\end{figure}

\begin{figure}[h]
\begin{center}
\includegraphics[scale=1.0]{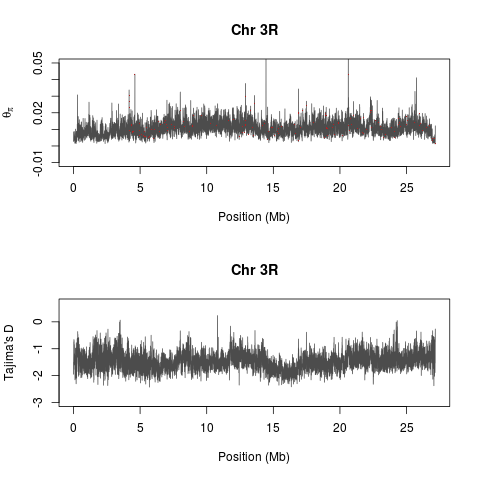}\end{center}
\caption{\label{ChromPlotsDsim3R} Nucleotide Diversity $\theta_{\pi}$ and \D {} chromosome 3R in \Dsim {} for 5kb windows with a 500 bp slide, showing only windows with 1 kb or more of sequence with coverage sufficient to call SNPS.  \D {} is negatively skewed, consistent with recent population expansion and mean diversity $\theta_{\pi}$ is low surrounding centromeric regions.  Locations and diversity values for duplications are marked in red.}

\end{figure}

\begin{figure}[h]
\begin{center}
\includegraphics[scale=1.0]{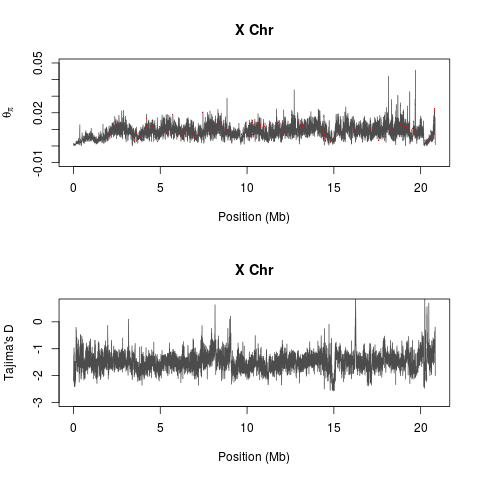}
\caption{\label{ChromPlotsDsimX} Nucleotide Diversity $\theta_{\pi}$ and \D {} for the X chromosome in  \Dsim {} for 5kb windows with a 500 bp slide, showing only windows with 1 kb or more of sequence with coverage sufficient to call SNPS.  \D {} is negatively skewed, consistent with recent population expansion and mean diversity $\theta_{\pi}$ is low surrounding centromeric regions. Locations and diversity values for duplications are marked in red.}
\end{center}
\end{figure}

\clearpage

\begin{figure}[h]
\begin{center}

\includegraphics[scale=0.80]{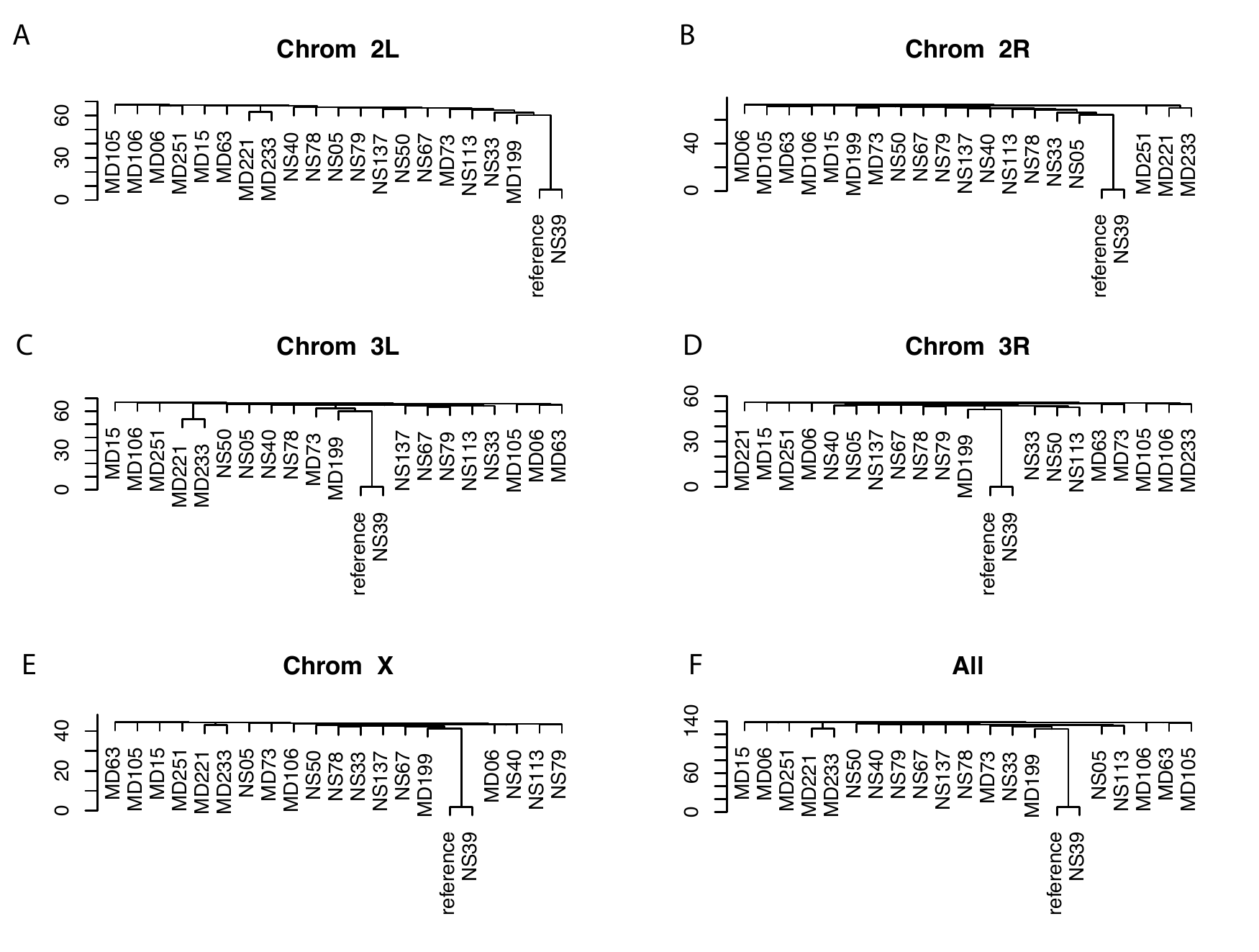}
\caption{\label{StructureSim}Hierarchical clustering of intronic SNP data for \Dsim {} shows little population structure. }
\end{center}

\end{figure}

\begin{figure}[h]
\begin{center}

\includegraphics[scale=0.80]{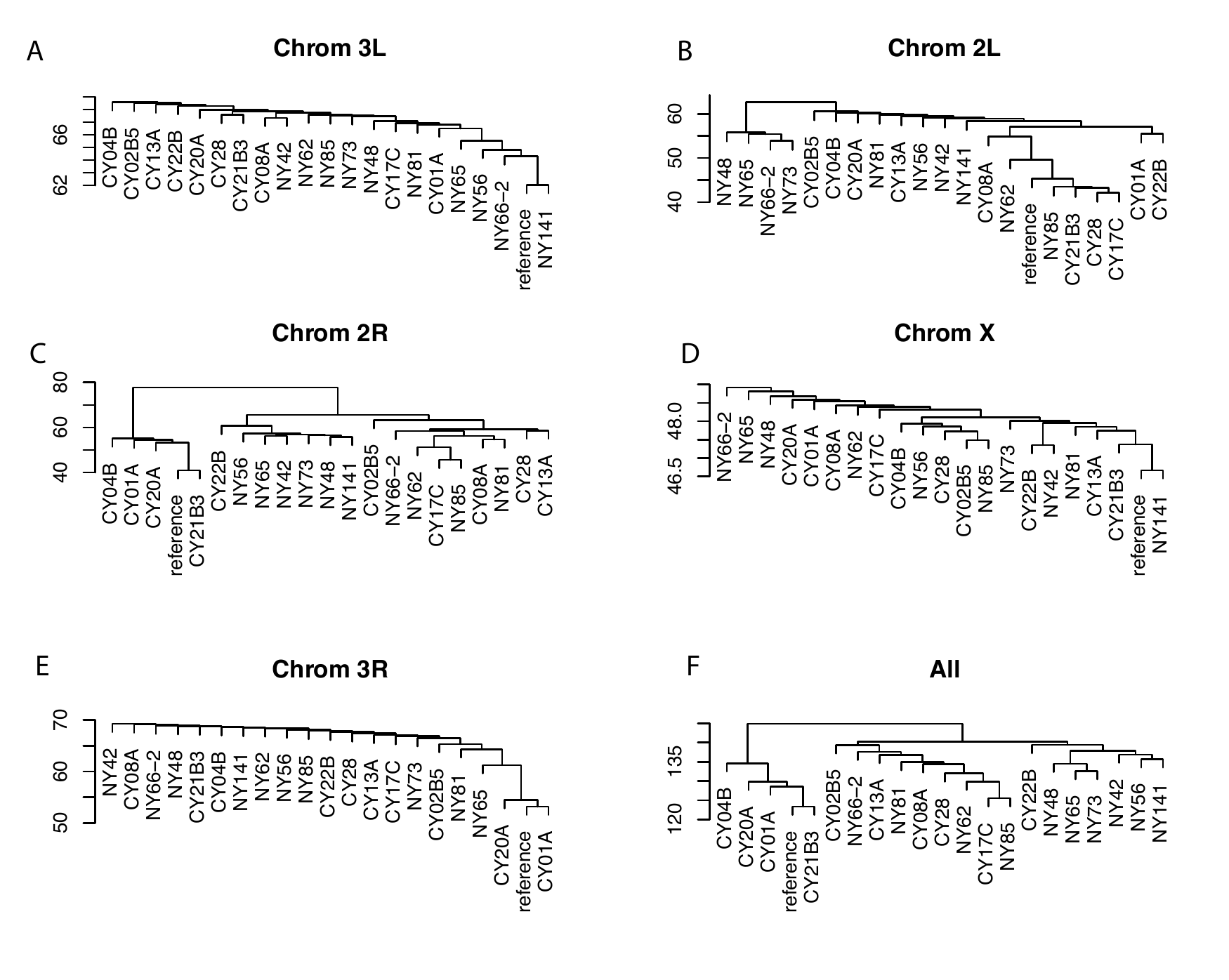}
\caption{\label{StructureYak}Hierarchical clustering of intronic SNP data for \Dyak {} shows population structure on chromosome 2L (B) and 2R (C), consistent with known inversions segregating on chromosome 2.  Samples do not cluster strictly with respect to geography (A,D-E), indicating widespread gene flow between geographic locations and a single admixed population. }
\end{center}

\end{figure}

\clearpage

\end{document}